\documentclass[prc,twocolumn,superscriptaddress,showpacs,twoside,floatfix]
{revtex4-1}

\usepackage{amssymb,epsfig}

\begin{document}
\title{The observation of long-range three-body Coulomb effects in
the decay of $^{16}$Ne}

\author{K.W.~Brown}
\affiliation{Departments of Chemistry and Physics, Washington University,
St.~Louis, Missouri 63130, USA.}
\author{R.J.~Charity}
\affiliation{Departments of Chemistry and Physics, Washington University,
St.~Louis, Missouri 63130, USA.}
\author{L.G.~Sobotka}
\affiliation{Departments of Chemistry and Physics, Washington University,
St.~Louis, Missouri 63130, USA.}
\author{Z.~Chajecki}
\affiliation{National Superconducting Cyclotron Laboratory and Department of
Physics and Astronomy, Michigan State University, East Lansing, Michigan 48824,
USA}
\author{L.V.~Grigorenko}
\affiliation{Flerov Laboratory of Nuclear Reactions, JINR, Dubna, RU-141980,
Russia}
\affiliation{Russian Research Center ``The Kurchatov Institute'', Kurchatov sq.\
1, RU-123182 Moscow, Russia}
\author{I.A.~Egorova}
\affiliation{Bogoliubov Laboratory of Theoretical Physics, JINR, Dubna, 141980,
Russia}
\author{Yu.L.~Parfenova}
\affiliation{Flerov Laboratory of Nuclear Reactions, JINR, Dubna, RU-141980
Russia}
\affiliation{Skobel'tsyn Institute of Nuclear Physics, Moscow State University,
119991 Moscow, Russia}
\author{M.V.~Zhukov}
\affiliation{Fundamental Physics, Chalmers University of Technology, S-41296
G\"{o}teborg, Sweden}
\author{S.~Bedoor}
\affiliation{Department of Physics, Western Michigan University, Kalamazoo,
Michigan 49008, USA.}
\author{W.W.~Buhro}
\affiliation{National Superconducting Cyclotron Laboratory and Department of
Physics and Astronomy, Michigan State University, East Lansing, Michigan 48824,
USA}
\author{J.M.~Elson}
\affiliation{Departments of Chemistry and Physics, Washington University,
St.~Louis, Missouri
63130, USA.}
\author{W.G.~Lynch}
\affiliation{National Superconducting Cyclotron Laboratory and Department of
Physics and Astronomy,  Michigan State University, East Lansing, Michigan 48824,
USA}
\author{J.~Manfredi}
\affiliation{National Superconducting Cyclotron Laboratory and Department of
Physics and Astronomy, Michigan State University, East Lansing, Michigan 48824,
USA}
\author{D.G. McNeel}
\affiliation{Department of Physics, Western Michigan University, Kalamazoo,
Michigan 49008, USA.}
\author{W.~Reviol}
\affiliation{Departments of Chemistry and Physics, Washington University,
St.~Louis, Missouri
63130, USA.}
\author{R.~Shane}
\affiliation{National Superconducting Cyclotron Laboratory and Department of
Physics and Astronomy,  Michigan State University, East Lansing, Michigan 48824,
USA}
\author{R.H.~Showalter}
\affiliation{National Superconducting Cyclotron Laboratory and Department of
Physics and Astronomy,  Michigan State University, East Lansing, Michigan 48824,
USA}
\author{M.B.~Tsang}
\affiliation{National Superconducting Cyclotron Laboratory and Department of
Physics and Astronomy,  Michigan State University, East Lansing, Michigan 48824,
USA}
\author{J.R.~Winkelbauer}
\affiliation{National Superconducting Cyclotron Laboratory and Department of
Physics and Astronomy,  Michigan State University, East Lansing, Michigan 48824,
USA}
\author {A.H.~Wuosmaa}
\affiliation{Department of Physics, Western Michigan University, Kalamazoo,
Michigan 49008, USA.}
\altaffiliation{Now at Department of Physics,University of Connecticut, Storrs,
CT 06269, USA}

\begin{abstract}
The interaction of an $E/A$=57.6-MeV $^{17}$Ne beam with a Be target was used
to populate levels in $^{16}$Ne following neutron knockout reactions.
The decay of $^{16}$Ne states into the three-body $^{14}$O+$p$+$p$ continuum
was observed in the High Resolution Array (HiRA). For the first time for a
2p emitter, correlations between the momenta of the three decay products
were measured  with sufficient resolution and statistics to allow for an
unambiguous demonstration of their dependence on the long-range nature
of the Coulomb interaction. Contrary to previous measurements, our measured 
limit $\Gamma<80$~keV for the intrinsic decay width of the ground 
state is not in contradiction with the small values (of the order of keV) 
predicted theoretically.
\end{abstract}


\maketitle


\textit{Introduction}
%
%
--- Two-proton ($2p$)  radioactivity \cite{Goldansky:1960} is the most recently
discovered type of radioactive decay. It is a facet of a broader three-body
decay phenomenon actively investigated within the last decade
\cite{Pfutzner:2012}. In binary decay, the correlations between the momenta of
the two decay products are entirely constrained by energy and momentum
conservation. In contrast for three-body decay, the corresponding correlations
are also sensitive to the internal nuclear structure of the decaying system and
the decay dynamics providing, in principle, another way to constrain this 
information from
experiment. In $2p$
decay, as the separation between the decay products becomes greater than the
range of the nuclear interaction, the subsequent modification of the initial
correlations is determined solely by the Coulomb interaction between the decay
products. As the range of the Coulomb force is infinite, its long-range
contribution to the correlations can be substantial, especially, in heavy $2p$
emitters.

Prompt $2p$  decay is a subset of a more general phenomenon of three-body
Coulomb decay (TBCD) which exists in mathematical physics (as a formal solution
of the $3 \rightarrow 3$ scattering of charged particles), in atomic physics (as
a solution of the $e \rightarrow 3e$ process), and in molecular physics (as
exotic molecules composed from three charged constituents)
\cite{Zaytsev:2013,McCurdy:2004,Hilico:2002,Kilic:2004,Madronero:2007,%
Ambrosio:2014}.  The theoretical treatment of TBCD is one of the oldest and most
complicated problems in physics because of the difficulty associated with the
boundary conditions due the the infinite range of the Coulomb force. The exact
analytical boundary conditions for this problem are unknown, but different
approximations to it have been tried. In nuclear physics, TBCD
has not attracted much attention, however the three-body Coulomb aspect of $2p$
decay will become increasingly important for heavier
prospective $2p$ emitters \cite{Olsen:2013}.

Detailed experimental studies of the correlations have been made for the
lightest $p$-shell $2p$ emitter $^{6}$Be \cite{Egorova:2012,Fomichev:2012} where
the Coulomb interactions are minute and their effects
are easily masked by the dynamics  of the nuclear interactions
\cite{Grigorenko:2012}. The Coulomb effects should be more prominent for the
heaviest observed $2p$ emitters, however these cases are limited
by poor statistics; e.g.\ the latest results for the $pf$-shell $2p$-emitters
$^{54}$Zn  \cite{Ascher:2011} and $^{45}$Fe \cite{Miernik:2007} are
based on just 7  and   75 events, respectively.
 Due to these limitations, previous $2p$ studies dedicated to the
long-range treatment of the three-body Coulomb interaction
\cite{Grigorenko:2010}, found consistency with the data, but no more.

The present work fills a gap between these previous studies by
measuring correlations in the $2p$ ground-state (g.s.) decay of the
$sd$-shell nucleus $^{16}$Ne where the Coulombic effects appear to be
strong enough to be observable. Known experimentally for several decades
\cite{Holt:1977}, $^{16}$Ne
has remained poorly investigated with just a few experimental studies
\cite{KeKelis:1978,Woodward:1983,Burleson:1980,Fohl:1997}. However, interest has
returned recently with the decay of $^{16}$Ne measured in relativistic
neutron-knockout reactions from a $^{17}$Ne beam \cite{Mukha:2010,Wamers:2014}.
We study the same reaction, but at an ``intermediate'' beam energy
and obtain data with better resolution and smaller statistical uncertainty.
Combined with state-of-the-art calculations, we find unambiguous evidence for
the role of the long-range Coulomb interactions in the measured correlations.

Apart from the Coulomb interactions,  predicted
correlations show sensitivity to the initial $2p$ configuration and nuclear
final-state interactions that are also evident in
$2n$ decay \cite{Kikuchi:2013,Grigorenko:2013,Hagino:2014}.
While there are indications of such sensitivities
in $2p$ data \cite{Pfutzner:2012}, the long-range Coulomb interactions must
first be determined accurately before the effects of structure and nuclear
final-state interactions can be better probed and properly accounted for.


\textit{Experiment}
%
%
--- A primary beam of $E/A$=170-MeV $^{20}$Ne, extracted from the Coupled
Cyclotron Facility at the National Superconducting Cyclotron Laboratory at
Michigan State University, bombarded a $^9$Be target.
The A1900 separator was used to select a secondary $^{17}$Ne beam  with a
momentum acceptance of $\pm 1.0\%$, an intensity
of $\sim 1.5 \times 10^5\text{ s}^{-1}$, and a purity of $11\%$ (the largest
component was $^{15}$O). This secondary beam impinged on a 1-mm-thick $^9$Be
target with an average of $E/A$=57.6~MeV in the target's center.

$^{16}$Ne decay products were detected in the High Resolution Array (HiRA) 
\cite{Wallace:2007} in an arrangement of fourteen $\Delta E-E$ [Si-CsI(Tl)] 
telescopes subtending zenith angles from $2^\circ$ to $13.9^\circ$ 
\cite{Charity:2011,Egorova:2012}. Energy calibrations were achieved using beams 
of 55 and 75~MeV protons and $E/A$=73 and 93~MeV $^{14}$O.


\textit{Theoretical model}
%
%
--- The model used in this work is similar to that applied to $^{16}$Ne in
\cite{Grigorenko:2002}, but, with improvements concerning basis convergence
\cite{Grigorenko:2009c}, TBCD \cite{Grigorenko:2010}, and the
reaction mechanism \cite{Egorova:2012}.
The three-body $^{14}$O+$p$+$p$ continuum of $^{16}$Ne is
described by the wave function (WF) $\Psi^{(+)}$ with the outgoing asymptotic
obtained by solving  the inhomogeneous three-body Schr\"odingier equation,
\[
(\hat{H}_3 - E_T)\Psi^{(+)} = \Phi_{\mathbf{q}},
\]
with approximate boundary conditions of the three-body Coulomb problem. The
three-body part of the model is based on the hyperspherical harmonics method
\cite{Grigorenko:2009c}. The differential cross section
is expressed via the flux $j$ induced by the WF $\Psi^{(+)}$ on the remote
surface $S$:
\begin{equation}
\frac{d \sigma}{d^3k_{^{14}\text{o}}d^3k_{p_1}d^3k_{p_2}} \sim j =  \left.
\langle
\Psi^{(+)} | \hat{j} | \Psi^{(+)} \rangle \right|_S \, .
\end{equation}
When comparing to the experimental data, the theoretical predictions  were used 
in Monte-Carlo (MC) simulations of the experiment 
\cite{Charity:2011,Egorova:2012} to take into account the apparatus
bias and resolution.

\begin{figure}
\begin{center}
\includegraphics[width=0.45\textwidth]{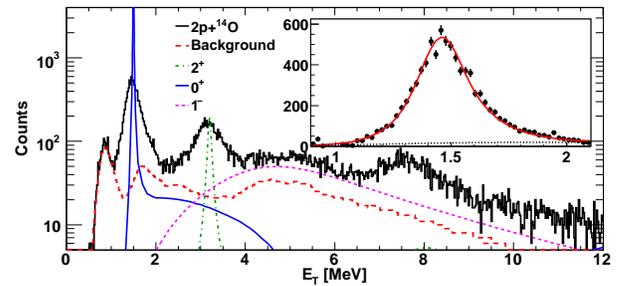}
\end{center}
\caption{(Color online) Experimental spectrum of $^{16}$Ne decay energy $E_T$
reconstructed from detected $^{14}$O+$p$+$p$ events. The dashed
histogram indicates the contamination from $^{15}$O+$p$+$p$ events. The smooth 
curves are
predictions (without detector resolution) for the indicated $^{16}$Ne states.
The inset compares the contamination-subtracted data to the
simulation of the g.s.\ peak for $\Gamma=0$, $f_{\text{tar}}=0.95$, where the 
dotted line is the
fitted background.}
\label{fig:exc-spec}
\end{figure}

The source function $\Phi_{\mathbf{q}}$ was approximated assuming the sudden 
removal of a neutron from the $^{15}$O core of $^{17}\text{Ne}_{\text{g.s.}}$,
\begin{equation}
\Phi_{\mathbf{q}} = \int d^3 r_n e^{i\mathbf{q r}_n} \langle
\Psi_{^{14}\text{\scriptsize
O}} | \Psi_{^{17}\text{\scriptsize Ne}} \rangle \, ,
\label{eq:sour}
\end{equation}
where $\mathbf{r}_n $ is the radius vector of the removed neutron. The 
$^{17}\text{Ne}_{\text{g.s.}}$ WF $\Psi_{^{17}\text{\scriptsize Ne}}$ was 
obtained in a three-body model of $^{15}$O+$p$+$p$ and broadly tested against 
various observables \cite{Grigorenko:2005}. Similar ideas had been applied to 
different reactions populating the three-body continuum of $^{6}$Be 
\cite{Egorova:2012,Grigorenko:2012,Fomichev:2012}. The $^{14}$O-$p$ potential 
sets were taken from \cite{Grigorenko:2002} which are consistent with a more 
recent experiment \cite{Goldberg:2004}, providing $1/2^+$ and $5/2^+$ states  at 
$E_r=1.45$ and 2.8 MeV, respectively consistent with the experimental properties 
of these states  in both $^{15}$F and $^{15}$C. We used the potential of 
\cite{Gogny:1970} for the $p$-$p$ channel.

The three-body Coulomb treatment in our model consists of two steps. (i) We are
able to impose approximate boundary conditions of TBCD on the
hypersphere of very large ($\rho_{\max} \lesssim 4000$ fm) hyperradius
 by diagonalizing the Coulomb interaction on the finite hyperspherical basis
\cite{Grigorenko:2001}. Within this limitation the procedure is exact, however
it breaks down at larger  hyperradii  as the accessible basis size become
insufficient. (ii) Classical trajectories are generated by a MC procedure at
the hyperradius $\rho_{\max}$ and propagated out to distances $\rho_{\text{ext}}
\gg \rho_{\max}$. The asymptotic momentum distributions are reconstructed from
the set of trajectories after the radial convergence is achieved. The accuracy
of this approach has been tested in calculations with simplified three-body
Hamiltonians allowing exact semi-analytical solutions \cite{Grigorenko:2010}.


\textit{Excitation spectrum}
%
%
--- The spectrum of the total decay energy $E_{T}$ constructed from the 
invariant mass of detected $^{14}$O+$p$+$p$ events is shown in 
Fig.~\ref{fig:exc-spec}. Due to a low-energy tail in the response function of 
the Si $\Delta E$ detectors, there is  leaking of a few $^{15}$O ions into the 
$^{14}$O gate in the $\Delta E-E$ spectrum. However, this contamination can be 
accurately modeled by taking detected $^{15}$O+$p$+$p$ events and analyzing them 
as $^{14}$O+$p$+$p$. The resulting spectrum (dashed histogram) was normalized to 
the $\sim1$-MeV peak associated with 2nd-excited state of $^{17}$Ne. All other 
peaks in the $^{16}$Ne spectrum are associated with $^{16}$Ne, with the g.s.\ 
peak at $E_T=1.466(20)$~MeV being the dominant feature. This decay energy is 
consistent with the value of 1.466(45)~MeV measured in \cite{Burleson:1980} and 
almost consistent with, but slightly larger than, other experimental values of 
1.34(8)~MeV \cite{KeKelis:1978}, 1.399(24)~MeV \cite{Woodward:1983}, and 
1.35(8)~MeV \cite{Mukha:2010}, 1.388(14)~MeV \cite {Wamers:2014}.

The predicted spectra in Fig.~\ref{fig:exc-spec} provide  guidance for possible 
spin-parity assignments of the other observed structures, suggesting that the 
previously known peaks \cite{Mukha:2010,Wamers:2014} at $E_T=3.16(2)$ and 
7.60(4)~MeV are both $2^+$ excited states. The broad structure at $E_T\sim 
5.0(5)$~MeV is well described as a $1^-$ ``soft'' excitation which is not a 
resonance, but a continuum mode, sensitive to the reaction mechanism 
\cite{Fomichev:2012}. In the mirror $^{16}$C system, there are also 
$J=2^{(\pm)}$, $3^{(+)}$, and $4^{+}$ contributions in this energy range, but 
for neutron-knockout from $p_{1/2}$, $p_{3/2}$, and $s_{1/2}$ orbitals in 
$^{17}$Ne, we should only expect \emph{strong} population for $0^+$, $2^+$, and 
$1^-$ configurations. We will concentrate on the g.s. for the remainder of this 
work ($1.27<E_{T}<1.72$~MeV) and all subsequent figures will show 
contamination-subtracted data.


\textit{Three-body energy-angular correlations}
%
%
--- The final state of a three-body decay can be completely
described by two parameters  \cite{Grigorenko:2009c}: an  energy
parameter $\varepsilon $ and an angle $\theta_{k}$ between the Jacobi momenta
$\mathbf{k}_{x}$, $\mathbf{k}_{y}$:
\begin{eqnarray}
\varepsilon = E_x/E_T \, ,\qquad \cos(\theta_k)=(\mathbf{k}_{x} \cdot
\mathbf{k}_{y}) /(k_x\,k_y) \, , \nonumber \\
{\bf k}_x  =  \frac{A_2 {\bf k}_1-A_1 {\bf k}_2 }{A_1+A_2} \, ,  \,\;
{\bf k}_y  =  \frac{A_3 ({\bf k}_1+{\bf k}_2)-(A_1+A_2) {\bf k}_3}
{A_1+A_2+A_3} , \nonumber \\
E_T =E_x+E_y=k^2_x/2M_x + k^2_y/2M_y ,  \qquad
\label{eq:corel-param}
\end{eqnarray}
where $M_x$ and $M_y$ are the reduced masses of the $X$ and $Y$ subsystems.
With the assignment $k_3 \rightarrow
k_{^{14}\text{O}}$, the correlations are obtained in the ``T'' Jacobi
system where $\varepsilon$ describes the relative energy $E_{pp}$ in the $p$-$p$
channel. For $k_3 \rightarrow k_{p}$, the correlations are obtained in one of
the
``Y'' Jacobi systems where $\varepsilon$ describes the relative energy
$E_{\text{core-}p}$ in the $^{14}$O-$p$ channel.

\begin{figure}
\center{
\includegraphics[width=0.48\textwidth]{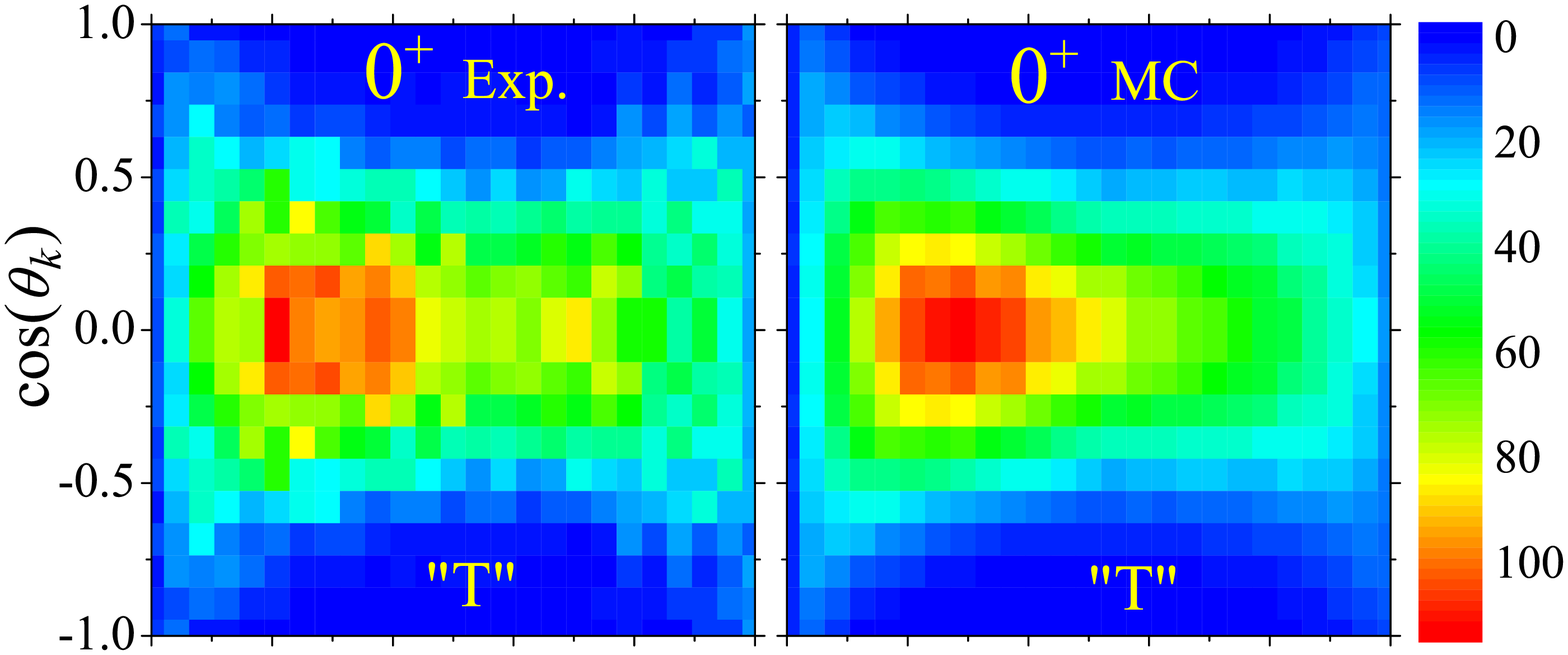}
\includegraphics[width=0.48\textwidth]{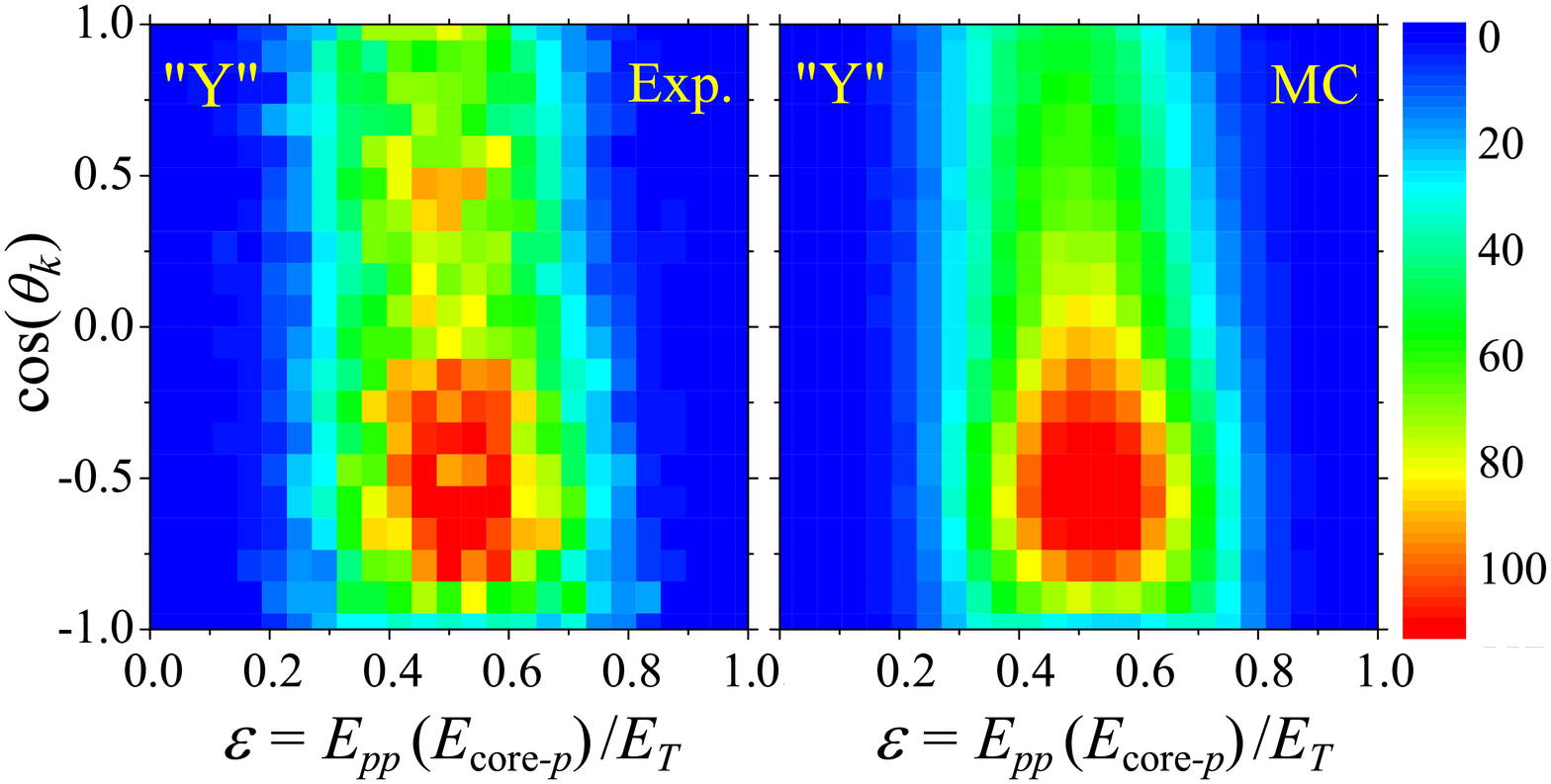}
}
\caption{(Color online) Energy-angular correlations for 
$^{16}\text{Ne}_{\text{g.s.}}$. Experimental and predicted (MC simulations) 
correlations for Jacobi ``T'' and ``Y'' systems are compared.}
\label{fig:complete}
\end{figure}

The experimental and predicted (MC simulations)  energy-angular
distributions, in both Jacobi representations are
compared in Fig.\ \ref{fig:complete} and found be similar.  More detailed
comparisons will be made with the projected energy distributions.

\begin{figure}[tb]
\begin{center}
\includegraphics[width=0.223\textwidth]{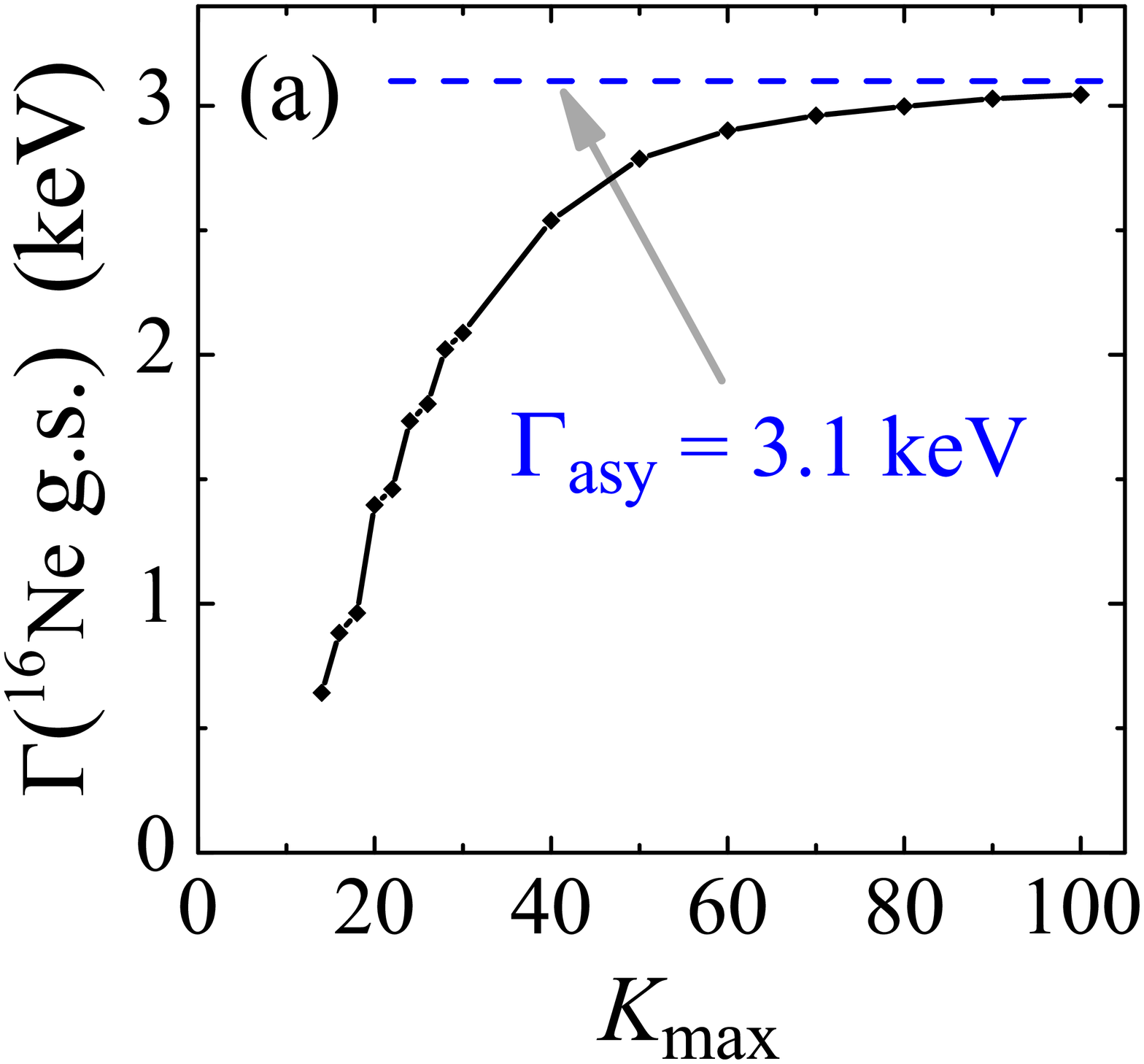}
\includegraphics[width=0.252\textwidth]{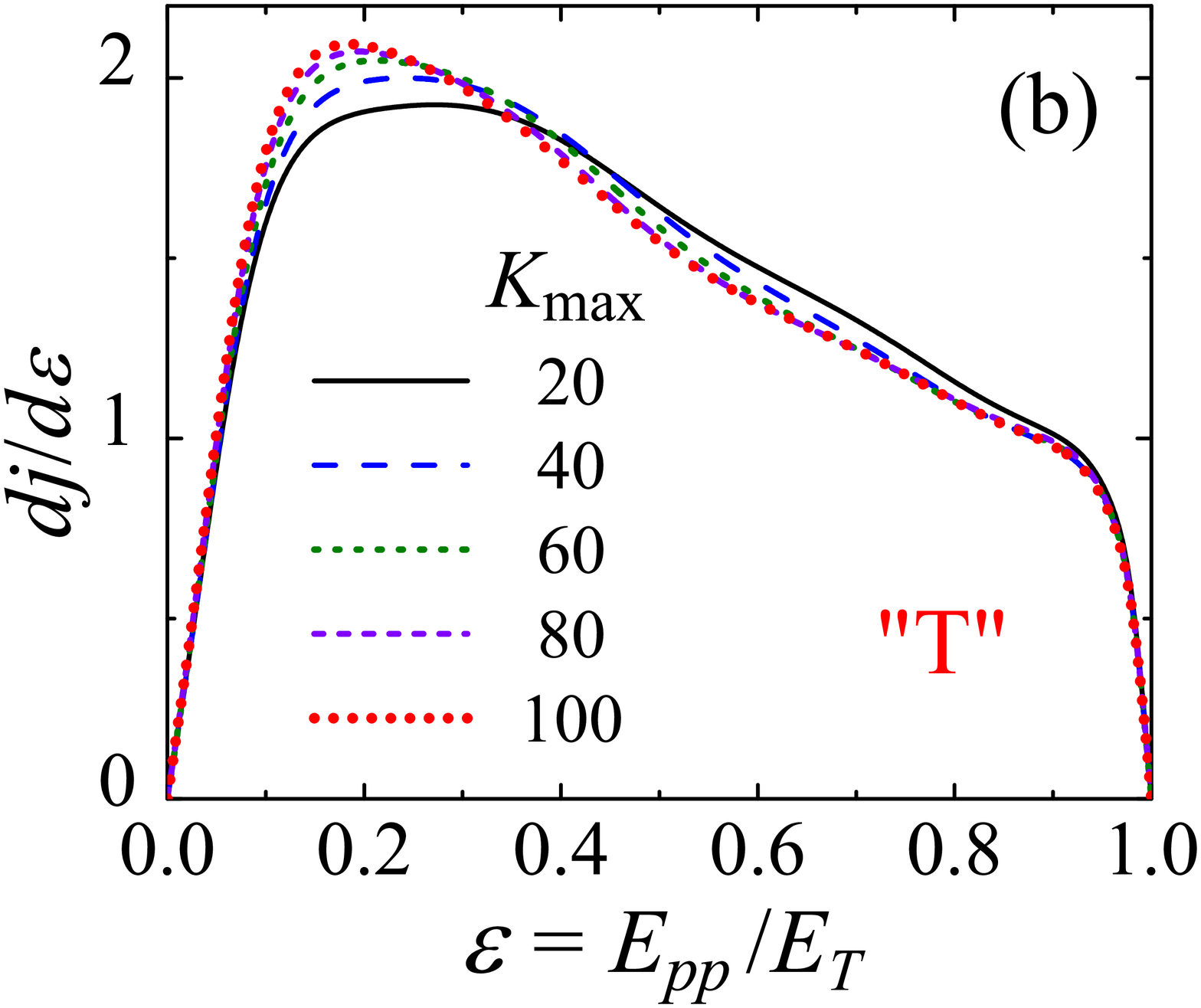}
\end{center}
\caption{(Color online) The convergence of the predicted (a) decay width and (b)
energy distribution in the ``T'' system  on $K_{\max}$ (maximum principal
quantum number of hyperspherical harmonics method). The asymptotic decay width 
of $^{16}$Ne assuming exponential $K_{\max}$ convergence is given in (a) by 
the dashed line.}
\label{fig:converg}
\end{figure}

\textit{The convergence of three-body calculations}
%
%
is quite slow for some observables \cite{Grigorenko:2009c,Grigorenko:2007}.
Figure \ref{fig:converg} demonstrates the convergence, with
increasing  $K_{\max}$ (maximum principle quantum number of the hyperspherical
harmonic method) for two observables for which the slowest convergence is 
expected. This work provides considerable improvement compared to the
calculations of \cite{Grigorenko:2002} which were limited by $K_{\max}=20$.

\begin{figure*}[tb]
\begin{center}
\includegraphics[width=0.33\textwidth]{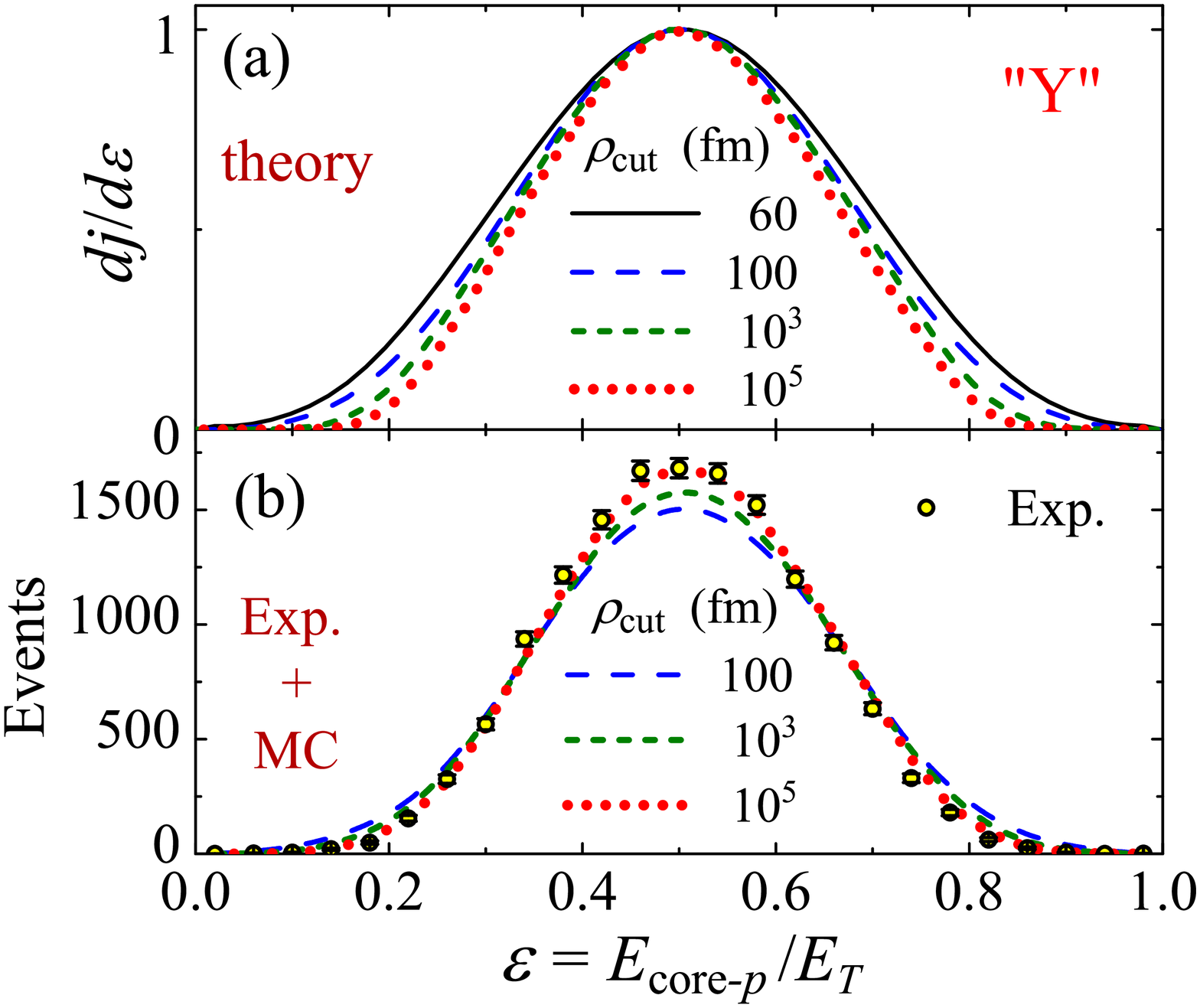}
\includegraphics[width=0.317\textwidth]{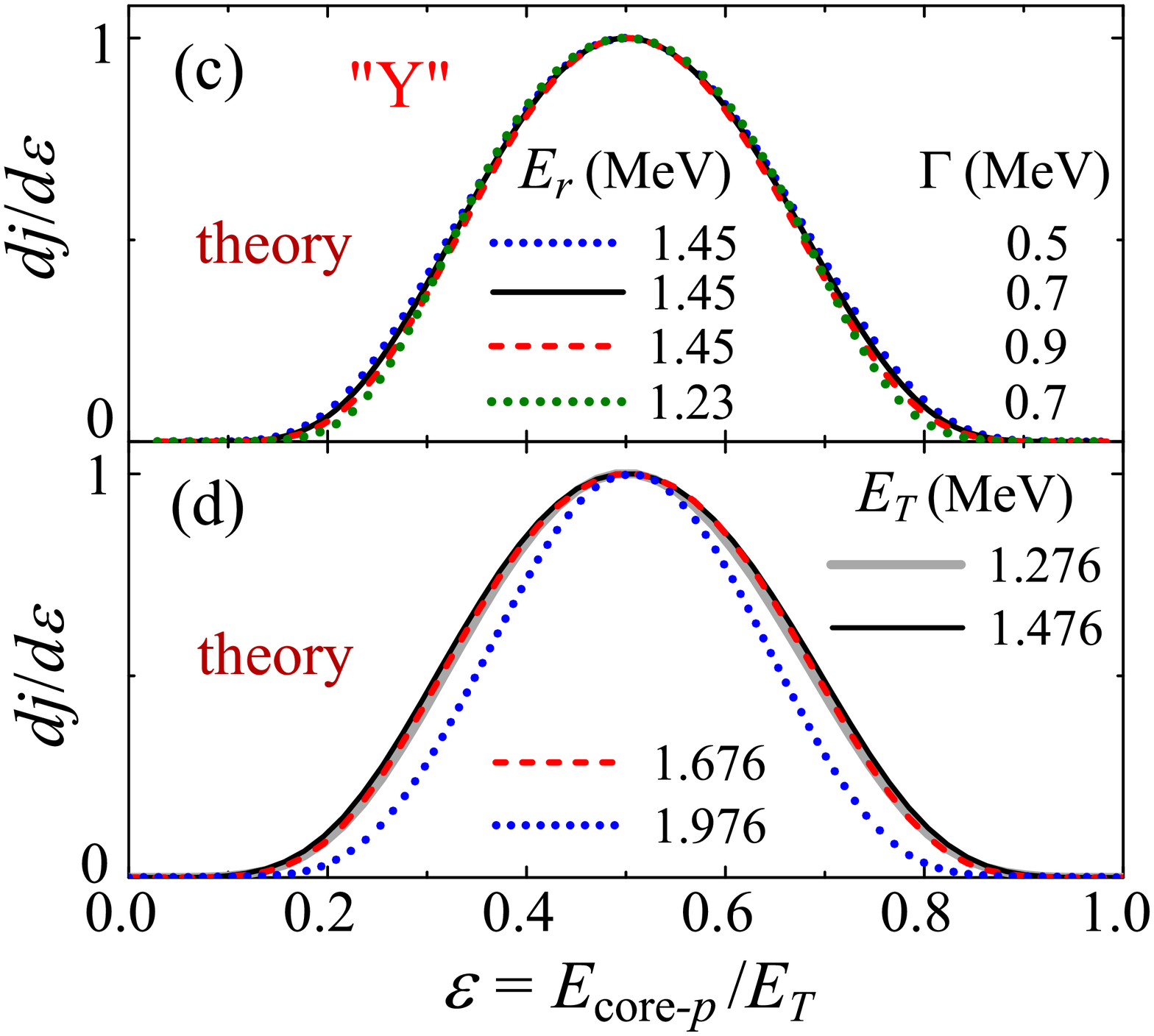}
\includegraphics[width=0.326\textwidth]{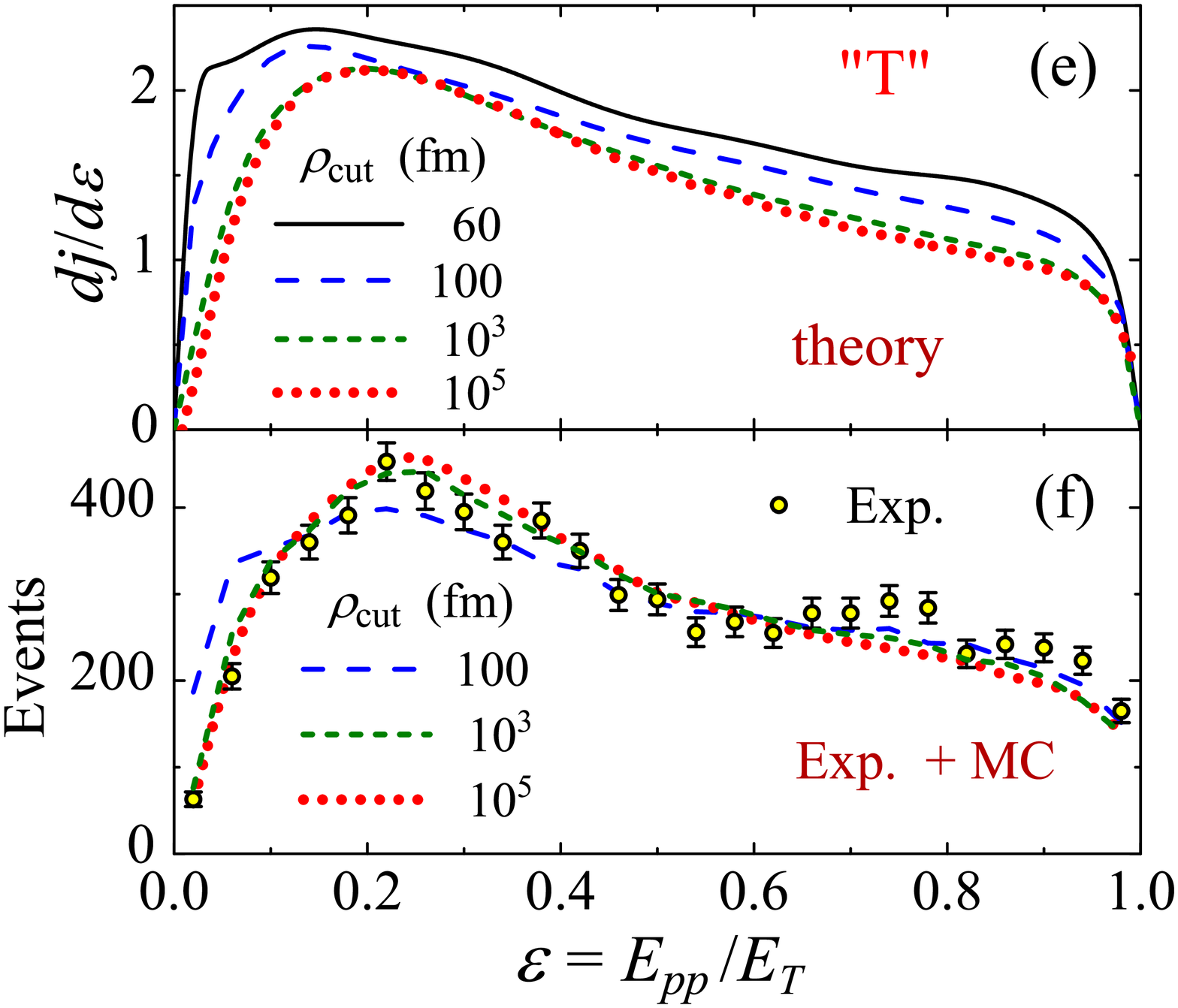}
\end{center}
\caption{(Color online) Panels (a)-(d) show energy distributions in the Jacobi
``Y'' system where (a) gives the sensitivity of the predictions to 
$\rho_{\text{cut}}$, (c) to the  $^{15}\text{F}_{\text{g.s.}}$  properties,
and (d) to the decay energy $E_{T}$. Panels (e), (f) show energy distributions 
in the Jacobi ``T'' system where (e) gives the sensitivity to 
$\rho_{\text{cut}}$. The theoretical predictions, after the detector bias is 
included via the MC simulations, are compared to the experimental data in (b) 
and (f) for the ``Y'' and ``T'' systems respectively. The normalization of the 
theoretical curves is arbitrary, while the MC results are normalized to the 
integral of the data.}
\label{fig:large}
\end{figure*}


\textit{$^{16}\text{Ne}_{\text{g.s.}}$ width}
%
%
--- The theoretical difficulty of reproducing the large experimental
g.s.\ widths measured for $^{12}$O and $^{16}$Ne has been pointed
out many times in the last 24 years
\cite{Korsheninnikov:1990,Azhari:1998,Grigorenko:2002,Barker:2003,
Fortune:2003}.
For $^{12}$O, this issue was resolved when a new  measurement \cite{Jager:2012}
gave a small upper bound. For
$^{16}$Ne, previous measurements of $\Gamma$=200(100)~keV
\cite{KeKelis:1978}, 110(40)~keV \cite{Woodward:1983}, and 82(15)~keV 
\cite{Wamers:2014} are large
compared to the theoretical predictions, e.g.\ 0.8 keV in
\cite{Grigorenko:2002}.

The experimental resolution is dominated by
the effects of multiple scattering and energy loss in the target.
Their magnitudes were fined tuned in the MC simulations
by reproducing  the experimental $^{15}$O+$p$+$p$ invariant-mass peak
associated with the narrow (predicted lifetime of 1.4 fs \cite{Chromik:2002})
 2nd-excited state in $^{17}$Ne
by scaling the target thickness from its known value by a factor 
$f_{\text{tar}}$.
The best fit is obtained with $f_{\text{tar}}^{\text{fit}}=0.95$ with
3-$\sigma$ limits of 0.91 and 1.00.
With $f_{\text{tar}}^{\text{fit}}$, we find that the simulated shape of the
$^{16}\text{Ne}_{\text{g.s.}}$ peak for $\Gamma=0$ is consistent with the data
[Fig.~\ref{fig:exc-spec} inset]. To obtain a limit for $\Gamma$, we
used a Breit-Wigner line shape in our simulations and find
a 3-$\sigma$ upper limit of $\Gamma<80$~keV with $f_{\text{tar}}=0.91$.
This limit is the first experimental result consistent with theoretical 
predictions of a small width [in the keV range, see, e.g.\ 
Fig.~\ref{fig:converg}(a)]. However, our limit is still considerably larger 
than the predictions, and on the other hand, it is still consistent 
with two of the previous experiments so even higher resolution
 measurements are needed to fully resolve this issue.


\textit{Evolution of energy distribution between core and proton.}
%
%
--- To investigate the long-range nature of TBCD, we studied the effect of 
terminating the Coulomb
interaction at some hyperradius $\rho_{\text{cut}}$. The energy distribution in 
the ``Y'' Jacobi system is largely
sensitive to just the TBCD and the global properties of the system ($E_T$,
charges, separation energies) \cite{Pfutzner:2012}. This makes it most suitable
for studying the $\rho_{\text{cut}}$ dependence [Fig.~\ref{fig:large}(a)].
Note the arbitrary normalization of the theoretical
curves, while the MC results are always normalized to the integral of the data.
The comparison with the data in Fig.~\ref{fig:large}(b)
demonstrates consistency with the theoretical calculations only if the
considered range of the Coulomb interaction far exceeds $10^3$~fm 
($\rho_{\text{cut}}= 10^{5}$~fm guarantees full convergence). This conclusion is 
only possible
due to the high quality of the present data. In contrast in  \cite{Wamers:2014}, 
where the experimental width of the g.s.\ peak
is almost twice as large and its integrated yield is $\sim 3$ times smaller, the
corresponding $\varepsilon$ distribution is broader with a FWHM of 0.41 compared
to our value of 0.33. This difference is similar to that obtained over
the range of $\rho_{\text{cut}}$ considered in Fig.~\ref{fig:large}(a)
demonstrating the need for high resolution to isolate these effects.

Our conclusions on TBCD are dependent on the stability of the
predicted correlations to the other inputs of the calculations.
Figure~\ref{fig:large}(d)  demonstrates the excellent stability of the core-$p$
energy distribution over a broad range ($\pm$200~keV) of  $E_{T}$ centered 
around $E_{T}$=1.476~MeV.
Indeed, in this range we have a maximum in the width for this distribution.
This maximum is expected as, below this range, the width must approach zero
in the limit of $E_{T}\rightarrow0$ \cite{Goldansky:1960} and, above this range,
we expect the width to have a minimum at $E_{T}\sim2E_{r}\sim$2.9~MeV,
 where $E_r$ is the $^{15}\text{F}_{\text{g.s.}}\rightarrow\text{core}+p$ decay energy. 
The predictions
of such a ``narrowing'' of the width at $E_{T}\sim2E_{r}$ \cite{Pfutzner:2012} were recently proven
experimentally \cite{Egorova:2012}. The curve for $E_T=1.976$ MeV is also
provided in Fig.~\ref{fig:large}(d) to show that a really large change in
 energy
is required to produce a significant modification of the $\varepsilon$ 
distribution.

The other important stability issue is with respect to the properties of
$^{15}\text{F}_{\text{g.s.}}$ for which there is \textit{no agreement} on its
centroid $E_r$ and width \cite{Fortune:2006}.
Figure~\ref{fig:large}(c) shows predicted $\varepsilon$ distributions based on
four different $^{14}$O+$p$ interactions which give the indicated 
$^{15}\text{F}_{\text{g.s.}}$ properties. Even if we use the data from 
\cite{Guo:2005}, which
differs the most from the other results ($E_r \sim 1.23$ MeV instead of
$E_r \sim 1.4-1.5$ MeV), no drastic effect is seen.


\textit{The evolution of energy distribution between two protons}
%
%
with $\rho_{\text{cut}}$ is shown in Fig.\ \ref{fig:large}(e).
This distribution has
greater sensitivity to the initial $2p$ configuration of the
decaying system \cite{Pfutzner:2012}. In addition,
the spin-singlet interaction
in the $p$-$p$ channel
provides the virtual state (``diproton'') which also
can affect the long-range behavior of the correlations
(see \cite{Kikuchi:2013,Grigorenko:2013,Hagino:2014} for the corresponding
effects in $2n$ decay). The
theoretical prediction for $\rho_{\text{cut}}=10^{5}$~fm in
Fig.\ \ref{fig:large}(f) reproduces
experimental data quite well, however, the sensitivity to $\rho_{\text{cut}}$
is diminished compared to the core-$p$ energy distribution.

\begin{figure}[tb]
\begin{center}
\includegraphics[width=0.36\textwidth]{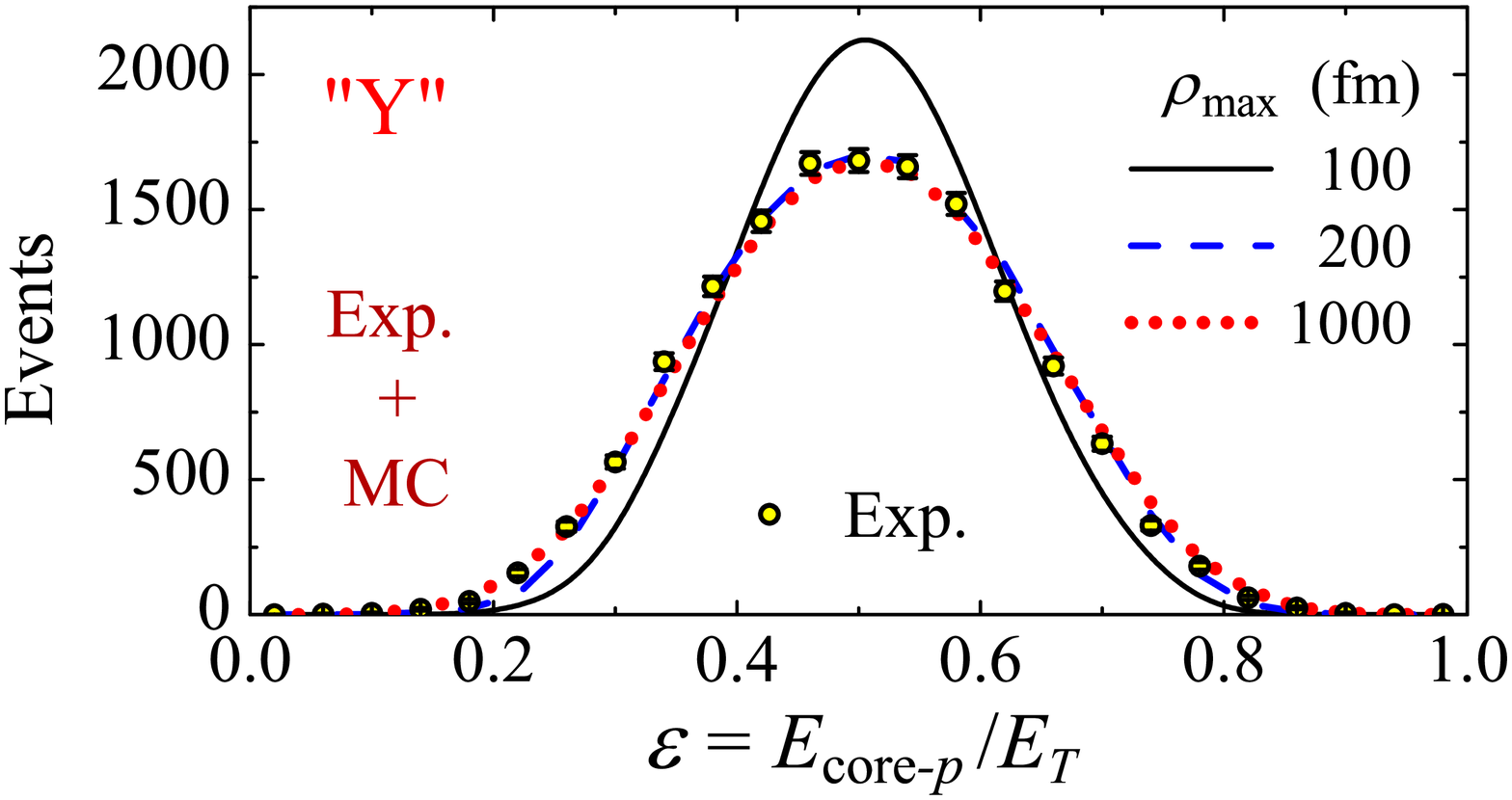}
\end{center}
\caption{The core-proton relative-energy distribution (``Y'' system)
obtained by classical extrapolation started from different $\rho_{\max}$ 
values.}
\label{fig:quasi-class}
\end{figure}


\textit{Limits on classical motion}
%
%
--- In our model the very long distances are achieved by classical 
extrapolation. This approximation has been studied using calculations with 
simplified Hamiltonians where it was demonstrated that the classical 
extrapolation provides stable results if the starting distance $\rho_{\max}$ 
exceeds some hundreds of fermis for $E_{T}\sim 1$~MeV \cite{Grigorenko:2010}. 
(e.g.\ $\sim 300$ fm for $^{19}\text{Mg}_{\text{g.s.}}$ decay where $E_{T}=0.75$ 
MeV). At such distances, the ratio of the Coulomb potential to the kinetic 
energy of fragments is of the order $10^{-2}$--$10^{-3}$. Figure 
\ref{fig:quasi-class} shows that for $^{16}\text{Ne}_{\text{g.s.}}$, the 
predictions are consistent with the data only if the conversion from quantum to 
classical dynamics is made at or above 200 fm.


\textit{Conclusions}
%
%
--- The continuum of $^{16}$Ne has been studied both experimentally and 
theoretically with emphasis on the ground state which decays by prompt 
two-proton emission. The measured decay correlations in this work were found to 
require a theoretical treatment in which the three-body Coulomb interaction is 
considered out to distances far beyond $10^{3}$ fm. Our theoretical treatment is 
now validated for use in interpreting the results of future studies of heavier 
two-proton decay with particular emphasis on extracting nuclear-structure 
information from correlation observables.

We extract a limit of  $\Gamma <80$~keV for the intrinsic decay width of the 
ground state, and while this is not inconsistant with some of the previous 
measurements, it is the first measurement consistent with the theoretical 
predictions. All conclusions of this work were only possible due to the high 
statistics and fidelity of the present measurements.

%
\textit{Acknowledgments}
%
%
--- This material is based upon work supported by the U.S.\ Department of 
Energy, Office of Science,  Office of
Nuclear Physics under Award numbers DE-FG02-87ER-40316 and DE-FG02-04ER41320 and 
the National Science Foundation under grants PHY-1102511 and PHY-9977707.
I.A.E.\ is supported by the Helmholtz Association under grant agreement
IK-RU-002 via FAIR-Russia Research Center and L.V.G.\ by the RFBR 14-02-00090
and Russian Ministry of Industry and Science NSh-932.2014.2 grants.
K.W.B.\ is supported by the National Science Foundation Graduate Research
Fellowship under grant No.\ DGE-1143954.



\begin{thebibliography}{42}%
\makeatletter
\providecommand \@ifxundefined [1]{%
 \@ifx{#1\undefined}
}%
\providecommand \@ifnum [1]{%
 \ifnum #1\expandafter \@firstoftwo
 \else \expandafter \@secondoftwo
 \fi
}%
\providecommand \@ifx [1]{%
 \ifx #1\expandafter \@firstoftwo
 \else \expandafter \@secondoftwo
 \fi
}%
\providecommand \natexlab [1]{#1}%
\providecommand \enquote  [1]{``#1''}%
\providecommand \bibnamefont  [1]{#1}%
\providecommand \bibfnamefont [1]{#1}%
\providecommand \citenamefont [1]{#1}%
\providecommand \href@noop [0]{\@secondoftwo}%
\providecommand \href [0]{\begingroup \@sanitize@url \@href}%
\providecommand \@href[1]{\@@startlink{#1}\@@href}%
\providecommand \@@href[1]{\endgroup#1\@@endlink}%
\providecommand \@sanitize@url [0]{\catcode `\\12\catcode `\$12\catcode
  `\&12\catcode `\#12\catcode `\^12\catcode `\_12\catcode `\%12\relax}%
\providecommand \@@startlink[1]{}%
\providecommand \@@endlink[0]{}%
\providecommand \url  [0]{\begingroup\@sanitize@url \@url }%
\providecommand \@url [1]{\endgroup\@href {#1}{\urlprefix }}%
\providecommand \urlprefix  [0]{URL }%
\providecommand \Eprint [0]{\href }%
\providecommand \doibase [0]{http://dx.doi.org/}%
\providecommand \selectlanguage [0]{\@gobble}%
\providecommand \bibinfo  [0]{\@secondoftwo}%
\providecommand \bibfield  [0]{\@secondoftwo}%
\providecommand \translation [1]{[#1]}%
\providecommand \BibitemOpen [0]{}%
\providecommand \bibitemStop [0]{}%
\providecommand \bibitemNoStop [0]{.\EOS\space}%
\providecommand \EOS [0]{\spacefactor3000\relax}%
\providecommand \BibitemShut  [1]{\csname bibitem#1\endcsname}%
\let\auto@bib@innerbib\@empty
\bibitem [{\citenamefont {Goldansky}(1960)}]{Goldansky:1960}%
  \BibitemOpen
  \bibfield  {author} {\bibinfo {author} {\bibfnamefont {V.~I.}\ \bibnamefont
  {Goldansky}},\ }\href@noop {} {\bibfield  {journal} {\bibinfo  {journal}
  {Nucl. Phys.}\ }\textbf {\bibinfo {volume} {19}},\ \bibinfo {pages} {482}
  (\bibinfo {year} {1960})}\BibitemShut {NoStop}%
\bibitem [{\citenamefont {Pf\"utzner}\ \emph {et~al.}(2012)\citenamefont
  {Pf\"utzner}, \citenamefont {Karny}, \citenamefont {Grigorenko},\ and\
  \citenamefont {Riisager}}]{Pfutzner:2012}%
  \BibitemOpen
  \bibfield  {author} {\bibinfo {author} {\bibfnamefont {M.}~\bibnamefont
  {Pf\"utzner}}, \bibinfo {author} {\bibfnamefont {M.}~\bibnamefont {Karny}},
  \bibinfo {author} {\bibfnamefont {L.~V.}\ \bibnamefont {Grigorenko}}, \ and\
  \bibinfo {author} {\bibfnamefont {K.}~\bibnamefont {Riisager}},\ }\href
  {\doibase 10.1103/RevModPhys.84.567} {\bibfield  {journal} {\bibinfo
  {journal} {Rev. Mod. Phys.}\ }\textbf {\bibinfo {volume} {84}},\ \bibinfo
  {pages} {567} (\bibinfo {year} {2012})}\BibitemShut {NoStop}%
\bibitem [{\citenamefont {Zaytsev}\ and\ \citenamefont
  {Gasaneo}(2013)}]{Zaytsev:2013}%
  \BibitemOpen
  \bibfield  {author} {\bibinfo {author} {\bibfnamefont {S.}~\bibnamefont
  {Zaytsev}}\ and\ \bibinfo {author} {\bibfnamefont {G.}~\bibnamefont
  {Gasaneo}},\ }\href@noop {} {\bibfield  {journal} {\bibinfo  {journal} {J.
  At. Mol. Sci.}\ }\textbf {\bibinfo {volume} {4}},\ \bibinfo {pages} {302}
  (\bibinfo {year} {2013})}\BibitemShut {NoStop}%
\bibitem [{\citenamefont {McCurdy}\ \emph {et~al.}(2004)\citenamefont
  {McCurdy}, \citenamefont {Baertschy},\ and\ \citenamefont
  {Rescigno}}]{McCurdy:2004}%
  \BibitemOpen
  \bibfield  {author} {\bibinfo {author} {\bibfnamefont {C.~W.}\ \bibnamefont
  {McCurdy}}, \bibinfo {author} {\bibfnamefont {M.}~\bibnamefont {Baertschy}},
  \ and\ \bibinfo {author} {\bibfnamefont {T.~N.}\ \bibnamefont {Rescigno}},\
  }\href@noop {} {\bibfield  {journal} {\bibinfo  {journal} {J. Phys. B: At.
  Mol. Opt. Phys.}\ }\textbf {\bibinfo {volume} {37}},\ \bibinfo {pages} {R137}
  (\bibinfo {year} {2004})}\BibitemShut {NoStop}%
\bibitem [{\citenamefont {Hilico}\ \emph {et~al.}(2002)\citenamefont {Hilico},
  \citenamefont {Gremaud}, \citenamefont {Jonckheere}, \citenamefont {Billy},\
  and\ \citenamefont {Delande}}]{Hilico:2002}%
  \BibitemOpen
  \bibfield  {author} {\bibinfo {author} {\bibfnamefont {L.}~\bibnamefont
  {Hilico}}, \bibinfo {author} {\bibfnamefont {B.}~\bibnamefont {Gremaud}},
  \bibinfo {author} {\bibfnamefont {T.}~\bibnamefont {Jonckheere}}, \bibinfo
  {author} {\bibfnamefont {N.}~\bibnamefont {Billy}}, \ and\ \bibinfo {author}
  {\bibfnamefont {D.}~\bibnamefont {Delande}},\ }\href@noop {} {\bibfield
  {journal} {\bibinfo  {journal} {Phys. Rev. A}\ }\textbf {\bibinfo {volume}
  {66}},\ \bibinfo {pages} {022101} (\bibinfo {year} {2002})}\BibitemShut
  {NoStop}%
\bibitem [{\citenamefont {Kilic}\ \emph {et~al.}(2004)\citenamefont {Kilic},
  \citenamefont {Karr},\ and\ \citenamefont {Hilico}}]{Kilic:2004}%
  \BibitemOpen
  \bibfield  {author} {\bibinfo {author} {\bibfnamefont {S.}~\bibnamefont
  {Kilic}}, \bibinfo {author} {\bibfnamefont {J.-P.}\ \bibnamefont {Karr}}, \
  and\ \bibinfo {author} {\bibfnamefont {L.}~\bibnamefont {Hilico}},\
  }\href@noop {} {\bibfield  {journal} {\bibinfo  {journal} {Phys. Rev. A}\
  }\textbf {\bibinfo {volume} {70}},\ \bibinfo {pages} {042506} (\bibinfo
  {year} {2004})}\BibitemShut {NoStop}%
\bibitem [{\citenamefont {Madronero}\ \emph {et~al.}(2007)\citenamefont
  {Madronero}, \citenamefont {Helico}, \citenamefont {Gremaud}, \citenamefont
  {Delande},\ and\ \citenamefont {Buchleitner}}]{Madronero:2007}%
  \BibitemOpen
  \bibfield  {author} {\bibinfo {author} {\bibfnamefont {J.}~\bibnamefont
  {Madronero}}, \bibinfo {author} {\bibfnamefont {L.}~\bibnamefont {Helico}},
  \bibinfo {author} {\bibfnamefont {B.}~\bibnamefont {Gremaud}}, \bibinfo
  {author} {\bibfnamefont {D.}~\bibnamefont {Delande}}, \ and\ \bibinfo
  {author} {\bibfnamefont {A.}~\bibnamefont {Buchleitner}},\ }\href@noop {}
  {\bibfield  {journal} {\bibinfo  {journal} {Math. Struct. in Comp. Science}\
  }\textbf {\bibinfo {volume} {17}},\ \bibinfo {pages} {225} (\bibinfo {year}
  {2007})}\BibitemShut {NoStop}%
\bibitem [{\citenamefont {Ambrosio}\ \emph {et~al.}(2014)\citenamefont
  {Ambrosio}, \citenamefont {Ancarani}, \citenamefont {Mitnik}, \citenamefont
  {Colavecchia},\ and\ \citenamefont {Gasaneo}}]{Ambrosio:2014}%
  \BibitemOpen
  \bibfield  {author} {\bibinfo {author} {\bibfnamefont {M.~J.}\ \bibnamefont
  {Ambrosio}}, \bibinfo {author} {\bibfnamefont {L.~U.}\ \bibnamefont
  {Ancarani}}, \bibinfo {author} {\bibfnamefont {D.~M.}\ \bibnamefont
  {Mitnik}}, \bibinfo {author} {\bibfnamefont {F.~D.}\ \bibnamefont
  {Colavecchia}}, \ and\ \bibinfo {author} {\bibfnamefont {G.}~\bibnamefont
  {Gasaneo}},\ }\href {\doibase 10.1007/s00601-014-0831-5} {\bibfield
  {journal} {\bibinfo  {journal} {Few-Body Systems}\ } (\bibinfo {year}
  {2014}),\ 10.1007/s00601-014-0831-5}\BibitemShut {NoStop}%
\bibitem [{\citenamefont {Olsen}\ \emph {et~al.}(2013)\citenamefont {Olsen},
  \citenamefont {Pf\"utzner}, \citenamefont {Birge}, \citenamefont {Brown},
  \citenamefont {Nazarewicz},\ and\ \citenamefont {Perhac}}]{Olsen:2013}%
  \BibitemOpen
  \bibfield  {author} {\bibinfo {author} {\bibfnamefont {E.}~\bibnamefont
  {Olsen}}, \bibinfo {author} {\bibfnamefont {M.}~\bibnamefont {Pf\"utzner}},
  \bibinfo {author} {\bibfnamefont {N.}~\bibnamefont {Birge}}, \bibinfo
  {author} {\bibfnamefont {M.}~\bibnamefont {Brown}}, \bibinfo {author}
  {\bibfnamefont {W.}~\bibnamefont {Nazarewicz}}, \ and\ \bibinfo {author}
  {\bibfnamefont {A.}~\bibnamefont {Perhac}},\ }\href {\doibase
  10.1103/PhysRevLett.110.222501} {\bibfield  {journal} {\bibinfo  {journal}
  {Phys. Rev. Lett.}\ }\textbf {\bibinfo {volume} {110}},\ \bibinfo {pages}
  {222501} (\bibinfo {year} {2013})}\BibitemShut {NoStop}%
\bibitem [{\citenamefont {Egorova}\ \emph {et~al.}(2012)\citenamefont
  {Egorova}, \citenamefont {Charity}, \citenamefont {Grigorenko}, \citenamefont
  {Chajecki}, \citenamefont {Coupland}, \citenamefont {Elson}, \citenamefont
  {Ghosh}, \citenamefont {Howard}, \citenamefont {Iwasaki}, \citenamefont
  {Kilburn}, \citenamefont {Lee}, \citenamefont {Lynch}, \citenamefont
  {Manfredi}, \citenamefont {Marley}, \citenamefont {Sanetullaev},
  \citenamefont {Shane}, \citenamefont {Shetty}, \citenamefont {Sobotka},
  \citenamefont {Tsang}, \citenamefont {Winkelbauer}, \citenamefont {Wuosmaa},
  \citenamefont {Youngs},\ and\ \citenamefont {Zhukov}}]{Egorova:2012}%
  \BibitemOpen
  \bibfield  {author} {\bibinfo {author} {\bibfnamefont {I.~A.}\ \bibnamefont
  {Egorova}}, \bibinfo {author} {\bibfnamefont {R.~J.}\ \bibnamefont
  {Charity}}, \bibinfo {author} {\bibfnamefont {L.~V.}\ \bibnamefont
  {Grigorenko}}, \bibinfo {author} {\bibfnamefont {Z.}~\bibnamefont
  {Chajecki}}, \bibinfo {author} {\bibfnamefont {D.}~\bibnamefont {Coupland}},
  \bibinfo {author} {\bibfnamefont {J.~M.}\ \bibnamefont {Elson}}, \bibinfo
  {author} {\bibfnamefont {T.~K.}\ \bibnamefont {Ghosh}}, \bibinfo {author}
  {\bibfnamefont {M.~E.}\ \bibnamefont {Howard}}, \bibinfo {author}
  {\bibfnamefont {H.}~\bibnamefont {Iwasaki}}, \bibinfo {author} {\bibfnamefont
  {M.}~\bibnamefont {Kilburn}}, \bibinfo {author} {\bibfnamefont
  {J.}~\bibnamefont {Lee}}, \bibinfo {author} {\bibfnamefont {W.~G.}\
  \bibnamefont {Lynch}}, \bibinfo {author} {\bibfnamefont {J.}~\bibnamefont
  {Manfredi}}, \bibinfo {author} {\bibfnamefont {S.~T.}\ \bibnamefont
  {Marley}}, \bibinfo {author} {\bibfnamefont {A.}~\bibnamefont {Sanetullaev}},
  \bibinfo {author} {\bibfnamefont {R.}~\bibnamefont {Shane}}, \bibinfo
  {author} {\bibfnamefont {D.~V.}\ \bibnamefont {Shetty}}, \bibinfo {author}
  {\bibfnamefont {L.~G.}\ \bibnamefont {Sobotka}}, \bibinfo {author}
  {\bibfnamefont {M.~B.}\ \bibnamefont {Tsang}}, \bibinfo {author}
  {\bibfnamefont {J.}~\bibnamefont {Winkelbauer}}, \bibinfo {author}
  {\bibfnamefont {A.~H.}\ \bibnamefont {Wuosmaa}}, \bibinfo {author}
  {\bibfnamefont {M.}~\bibnamefont {Youngs}}, \ and\ \bibinfo {author}
  {\bibfnamefont {M.~V.}\ \bibnamefont {Zhukov}},\ }\href {\doibase
  10.1103/PhysRevLett.109.202502} {\bibfield  {journal} {\bibinfo  {journal}
  {Phys. Rev. Lett.}\ }\textbf {\bibinfo {volume} {109}},\ \bibinfo {pages}
  {202502} (\bibinfo {year} {2012})}\BibitemShut {NoStop}%
\bibitem [{\citenamefont {Fomichev}\ \emph {et~al.}(2012)\citenamefont
  {Fomichev}, \citenamefont {Chudoba}, \citenamefont {Egorova}, \citenamefont
  {Ershov}, \citenamefont {Golovkov}, \citenamefont {Gorshkov}, \citenamefont
  {Gorshkov}, \citenamefont {Grigorenko}, \citenamefont {Kamiński},
  \citenamefont {Krupko}, \citenamefont {Mukha}, \citenamefont {Parfenova},
  \citenamefont {Sidorchuk}, \citenamefont {Slepnev}, \citenamefont
  {Standyło}, \citenamefont {Stepantsov}, \citenamefont {Ter-Akopian},
  \citenamefont {Wolski},\ and\ \citenamefont {Zhukov}}]{Fomichev:2012}%
  \BibitemOpen
  \bibfield  {author} {\bibinfo {author} {\bibfnamefont {A.}~\bibnamefont
  {Fomichev}}, \bibinfo {author} {\bibfnamefont {V.}~\bibnamefont {Chudoba}},
  \bibinfo {author} {\bibfnamefont {I.}~\bibnamefont {Egorova}}, \bibinfo
  {author} {\bibfnamefont {S.}~\bibnamefont {Ershov}}, \bibinfo {author}
  {\bibfnamefont {M.}~\bibnamefont {Golovkov}}, \bibinfo {author}
  {\bibfnamefont {A.}~\bibnamefont {Gorshkov}}, \bibinfo {author}
  {\bibfnamefont {V.}~\bibnamefont {Gorshkov}}, \bibinfo {author}
  {\bibfnamefont {L.}~\bibnamefont {Grigorenko}}, \bibinfo {author}
  {\bibfnamefont {G.}~\bibnamefont {Kamiński}}, \bibinfo {author}
  {\bibfnamefont {S.}~\bibnamefont {Krupko}}, \bibinfo {author} {\bibfnamefont
  {I.}~\bibnamefont {Mukha}}, \bibinfo {author} {\bibfnamefont
  {Y.}~\bibnamefont {Parfenova}}, \bibinfo {author} {\bibfnamefont
  {S.}~\bibnamefont {Sidorchuk}}, \bibinfo {author} {\bibfnamefont
  {R.}~\bibnamefont {Slepnev}}, \bibinfo {author} {\bibfnamefont
  {L.}~\bibnamefont {Standyło}}, \bibinfo {author} {\bibfnamefont
  {S.}~\bibnamefont {Stepantsov}}, \bibinfo {author} {\bibfnamefont
  {G.}~\bibnamefont {Ter-Akopian}}, \bibinfo {author} {\bibfnamefont
  {R.}~\bibnamefont {Wolski}}, \ and\ \bibinfo {author} {\bibfnamefont
  {M.}~\bibnamefont {Zhukov}},\ }\href {\doibase
  http://dx.doi.org/10.1016/j.physletb.2012.01.004} {\bibfield  {journal}
  {\bibinfo  {journal} {Physics Letters B}\ }\textbf {\bibinfo {volume}
  {708}},\ \bibinfo {pages} {6 } (\bibinfo {year} {2012})}\BibitemShut
  {NoStop}%
\bibitem [{\citenamefont {Grigorenko}\ \emph {et~al.}(2012)\citenamefont
  {Grigorenko}, \citenamefont {Egorova}, \citenamefont {Charity},\ and\
  \citenamefont {Zhukov}}]{Grigorenko:2012}%
  \BibitemOpen
  \bibfield  {author} {\bibinfo {author} {\bibfnamefont {L.~V.}\ \bibnamefont
  {Grigorenko}}, \bibinfo {author} {\bibfnamefont {I.~A.}\ \bibnamefont
  {Egorova}}, \bibinfo {author} {\bibfnamefont {R.~J.}\ \bibnamefont
  {Charity}}, \ and\ \bibinfo {author} {\bibfnamefont {M.~V.}\ \bibnamefont
  {Zhukov}},\ }\href {\doibase 10.1103/PhysRevC.86.061602} {\bibfield
  {journal} {\bibinfo  {journal} {Phys. Rev. C}\ }\textbf {\bibinfo {volume}
  {86}},\ \bibinfo {pages} {061602} (\bibinfo {year} {2012})}\BibitemShut
  {NoStop}%
\bibitem [{\citenamefont {Ascher}\ \emph {et~al.}(2011)\citenamefont {Ascher},
  \citenamefont {Audirac}, \citenamefont {Adimi}, \citenamefont {Blank},
  \citenamefont {Borcea}, \citenamefont {Brown}, \citenamefont {Companis},
  \citenamefont {Delalee}, \citenamefont {Demonchy}, \citenamefont
  {de~Oliveira~Santos}, \citenamefont {Giovinazzo}, \citenamefont {Gr\'evy},
  \citenamefont {Grigorenko}, \citenamefont {Kurtukian-Nieto}, \citenamefont
  {Leblanc}, \citenamefont {Pedroza}, \citenamefont {Perrot}, \citenamefont
  {Pibernat}, \citenamefont {Serani}, \citenamefont {Srivastava},\ and\
  \citenamefont {Thomas}}]{Ascher:2011}%
  \BibitemOpen
  \bibfield  {author} {\bibinfo {author} {\bibfnamefont {P.}~\bibnamefont
  {Ascher}}, \bibinfo {author} {\bibfnamefont {L.}~\bibnamefont {Audirac}},
  \bibinfo {author} {\bibfnamefont {N.}~\bibnamefont {Adimi}}, \bibinfo
  {author} {\bibfnamefont {B.}~\bibnamefont {Blank}}, \bibinfo {author}
  {\bibfnamefont {C.}~\bibnamefont {Borcea}}, \bibinfo {author} {\bibfnamefont
  {B.~A.}\ \bibnamefont {Brown}}, \bibinfo {author} {\bibfnamefont
  {I.}~\bibnamefont {Companis}}, \bibinfo {author} {\bibfnamefont
  {F.}~\bibnamefont {Delalee}}, \bibinfo {author} {\bibfnamefont {C.~E.}\
  \bibnamefont {Demonchy}}, \bibinfo {author} {\bibfnamefont {F.}~\bibnamefont
  {de~Oliveira~Santos}}, \bibinfo {author} {\bibfnamefont {J.}~\bibnamefont
  {Giovinazzo}}, \bibinfo {author} {\bibfnamefont {S.}~\bibnamefont {Gr\'evy}},
  \bibinfo {author} {\bibfnamefont {L.~V.}\ \bibnamefont {Grigorenko}},
  \bibinfo {author} {\bibfnamefont {T.}~\bibnamefont {Kurtukian-Nieto}},
  \bibinfo {author} {\bibfnamefont {S.}~\bibnamefont {Leblanc}}, \bibinfo
  {author} {\bibfnamefont {J.-L.}\ \bibnamefont {Pedroza}}, \bibinfo {author}
  {\bibfnamefont {L.}~\bibnamefont {Perrot}}, \bibinfo {author} {\bibfnamefont
  {J.}~\bibnamefont {Pibernat}}, \bibinfo {author} {\bibfnamefont
  {L.}~\bibnamefont {Serani}}, \bibinfo {author} {\bibfnamefont {P.~C.}\
  \bibnamefont {Srivastava}}, \ and\ \bibinfo {author} {\bibfnamefont {J.-C.}\
  \bibnamefont {Thomas}},\ }\href {\doibase 10.1103/PhysRevLett.107.102502}
  {\bibfield  {journal} {\bibinfo  {journal} {Phys. Rev. Lett.}\ }\textbf
  {\bibinfo {volume} {107}},\ \bibinfo {pages} {102502} (\bibinfo {year}
  {2011})}\BibitemShut {NoStop}%
\bibitem [{\citenamefont {Miernik}\ \emph {et~al.}(2007)\citenamefont
  {Miernik}, \citenamefont {Dominik}, \citenamefont {Janas}, \citenamefont
  {Pf\"utzner}, \citenamefont {Grigorenko}, \citenamefont {Bingham},
  \citenamefont {Czyrkowski}, \citenamefont {Cwiok}, \citenamefont {Darby},
  \citenamefont {Dabrowski}, \citenamefont {Ginter}, \citenamefont {Grzywacz},
  \citenamefont {Karny}, \citenamefont {Korgul}, \citenamefont {Kusmierz},
  \citenamefont {Liddick}, \citenamefont {Rajabali}, \citenamefont
  {Rykaczewski},\ and\ \citenamefont {Stolz}}]{Miernik:2007}%
  \BibitemOpen
  \bibfield  {author} {\bibinfo {author} {\bibfnamefont {K.}~\bibnamefont
  {Miernik}}, \bibinfo {author} {\bibfnamefont {W.}~\bibnamefont {Dominik}},
  \bibinfo {author} {\bibfnamefont {Z.}~\bibnamefont {Janas}}, \bibinfo
  {author} {\bibfnamefont {M.}~\bibnamefont {Pf\"utzner}}, \bibinfo {author}
  {\bibfnamefont {L.}~\bibnamefont {Grigorenko}}, \bibinfo {author}
  {\bibfnamefont {C.~R.}\ \bibnamefont {Bingham}}, \bibinfo {author}
  {\bibfnamefont {H.}~\bibnamefont {Czyrkowski}}, \bibinfo {author}
  {\bibfnamefont {M.}~\bibnamefont {Cwiok}}, \bibinfo {author} {\bibfnamefont
  {I.~G.}\ \bibnamefont {Darby}}, \bibinfo {author} {\bibfnamefont
  {R.}~\bibnamefont {Dabrowski}}, \bibinfo {author} {\bibfnamefont
  {T.}~\bibnamefont {Ginter}}, \bibinfo {author} {\bibfnamefont
  {R.}~\bibnamefont {Grzywacz}}, \bibinfo {author} {\bibfnamefont
  {M.}~\bibnamefont {Karny}}, \bibinfo {author} {\bibfnamefont
  {A.}~\bibnamefont {Korgul}}, \bibinfo {author} {\bibfnamefont
  {W.}~\bibnamefont {Kusmierz}}, \bibinfo {author} {\bibfnamefont {S.~N.}\
  \bibnamefont {Liddick}}, \bibinfo {author} {\bibfnamefont {M.}~\bibnamefont
  {Rajabali}}, \bibinfo {author} {\bibfnamefont {K.}~\bibnamefont
  {Rykaczewski}}, \ and\ \bibinfo {author} {\bibfnamefont {A.}~\bibnamefont
  {Stolz}},\ }\href@noop {} {\bibfield  {journal} {\bibinfo  {journal} {Phys.
  Rev. Lett.}\ }\textbf {\bibinfo {volume} {99}},\ \bibinfo {pages} {192501}
  (\bibinfo {year} {2007})}\BibitemShut {NoStop}%
\bibitem [{\citenamefont {Grigorenko}\ \emph {et~al.}(2010)\citenamefont
  {Grigorenko}, \citenamefont {Egorova}, \citenamefont {Zhukov}, \citenamefont
  {Charity},\ and\ \citenamefont {Miernik}}]{Grigorenko:2010}%
  \BibitemOpen
  \bibfield  {author} {\bibinfo {author} {\bibfnamefont {L.~V.}\ \bibnamefont
  {Grigorenko}}, \bibinfo {author} {\bibfnamefont {I.~A.}\ \bibnamefont
  {Egorova}}, \bibinfo {author} {\bibfnamefont {M.~V.}\ \bibnamefont {Zhukov}},
  \bibinfo {author} {\bibfnamefont {R.~J.}\ \bibnamefont {Charity}}, \ and\
  \bibinfo {author} {\bibfnamefont {K.}~\bibnamefont {Miernik}},\ }\href
  {\doibase 10.1103/PhysRevC.82.014615} {\bibfield  {journal} {\bibinfo
  {journal} {Phys. Rev. C}\ }\textbf {\bibinfo {volume} {82}},\ \bibinfo
  {pages} {014615} (\bibinfo {year} {2010})}\BibitemShut {NoStop}%
\bibitem [{\citenamefont {Holt}\ \emph {et~al.}(1977)\citenamefont {Holt},
  \citenamefont {Zeidman}, \citenamefont {Malbrough}, \citenamefont {Marks},
  \citenamefont {Preedom}, \citenamefont {Baker}, \citenamefont {Burman},
  \citenamefont {Cooper}, \citenamefont {Heffner}, \citenamefont {Lee},
  \citenamefont {Redwine},\ and\ \citenamefont {Spencer}}]{Holt:1977}%
  \BibitemOpen
  \bibfield  {author} {\bibinfo {author} {\bibfnamefont {R.}~\bibnamefont
  {Holt}}, \bibinfo {author} {\bibfnamefont {B.}~\bibnamefont {Zeidman}},
  \bibinfo {author} {\bibfnamefont {D.}~\bibnamefont {Malbrough}}, \bibinfo
  {author} {\bibfnamefont {T.}~\bibnamefont {Marks}}, \bibinfo {author}
  {\bibfnamefont {B.}~\bibnamefont {Preedom}}, \bibinfo {author} {\bibfnamefont
  {M.}~\bibnamefont {Baker}}, \bibinfo {author} {\bibfnamefont
  {R.}~\bibnamefont {Burman}}, \bibinfo {author} {\bibfnamefont
  {M.}~\bibnamefont {Cooper}}, \bibinfo {author} {\bibfnamefont
  {R.}~\bibnamefont {Heffner}}, \bibinfo {author} {\bibfnamefont
  {D.}~\bibnamefont {Lee}}, \bibinfo {author} {\bibfnamefont {R.}~\bibnamefont
  {Redwine}}, \ and\ \bibinfo {author} {\bibfnamefont {J.}~\bibnamefont
  {Spencer}},\ }\href {\doibase http://dx.doi.org/10.1016/0370-2693(77)90131-9}
  {\bibfield  {journal} {\bibinfo  {journal} {Physics Letters B}\ }\textbf
  {\bibinfo {volume} {69}},\ \bibinfo {pages} {55 } (\bibinfo {year}
  {1977})}\BibitemShut {NoStop}%
\bibitem [{\citenamefont {KeKelis}\ \emph {et~al.}(1978)\citenamefont
  {KeKelis}, \citenamefont {Zisman}, \citenamefont {Scott}, \citenamefont
  {Jahn}, \citenamefont {Vieira}, \citenamefont {Cerny},\ and\ \citenamefont
  {Ajzenberg-Selove}}]{KeKelis:1978}%
  \BibitemOpen
  \bibfield  {author} {\bibinfo {author} {\bibfnamefont {G.~J.}\ \bibnamefont
  {KeKelis}}, \bibinfo {author} {\bibfnamefont {M.~S.}\ \bibnamefont {Zisman}},
  \bibinfo {author} {\bibfnamefont {D.~K.}\ \bibnamefont {Scott}}, \bibinfo
  {author} {\bibfnamefont {R.}~\bibnamefont {Jahn}}, \bibinfo {author}
  {\bibfnamefont {D.~J.}\ \bibnamefont {Vieira}}, \bibinfo {author}
  {\bibfnamefont {J.}~\bibnamefont {Cerny}}, \ and\ \bibinfo {author}
  {\bibfnamefont {F.}~\bibnamefont {Ajzenberg-Selove}},\ }\href {\doibase
  10.1103/PhysRevC.17.1929} {\bibfield  {journal} {\bibinfo  {journal} {Phys.
  Rev. C}\ }\textbf {\bibinfo {volume} {17}},\ \bibinfo {pages} {1929}
  (\bibinfo {year} {1978})}\BibitemShut {NoStop}%
\bibitem [{\citenamefont {Woodward}\ \emph {et~al.}(1983)\citenamefont
  {Woodward}, \citenamefont {Tribble},\ and\ \citenamefont
  {Tanner}}]{Woodward:1983}%
  \BibitemOpen
  \bibfield  {author} {\bibinfo {author} {\bibfnamefont {C.~J.}\ \bibnamefont
  {Woodward}}, \bibinfo {author} {\bibfnamefont {R.~E.}\ \bibnamefont
  {Tribble}}, \ and\ \bibinfo {author} {\bibfnamefont {D.~M.}\ \bibnamefont
  {Tanner}},\ }\href {\doibase 10.1103/PhysRevC.27.27} {\bibfield  {journal}
  {\bibinfo  {journal} {Phys. Rev. C}\ }\textbf {\bibinfo {volume} {27}},\
  \bibinfo {pages} {27} (\bibinfo {year} {1983})}\BibitemShut {NoStop}%
\bibitem [{\citenamefont {Burleson}\ \emph {et~al.}(1980)\citenamefont
  {Burleson}, \citenamefont {Blanpied}, \citenamefont {Daw}, \citenamefont
  {Viescas}, \citenamefont {Morris}, \citenamefont {Thiessen}, \citenamefont
  {Greene}, \citenamefont {Braithwaite}, \citenamefont {Cottingame},
  \citenamefont {Holtkamp}, \citenamefont {Moore},\ and\ \citenamefont
  {Moore}}]{Burleson:1980}%
  \BibitemOpen
  \bibfield  {author} {\bibinfo {author} {\bibfnamefont {G.~R.}\ \bibnamefont
  {Burleson}}, \bibinfo {author} {\bibfnamefont {G.~S.}\ \bibnamefont
  {Blanpied}}, \bibinfo {author} {\bibfnamefont {G.~H.}\ \bibnamefont {Daw}},
  \bibinfo {author} {\bibfnamefont {A.~J.}\ \bibnamefont {Viescas}}, \bibinfo
  {author} {\bibfnamefont {C.~L.}\ \bibnamefont {Morris}}, \bibinfo {author}
  {\bibfnamefont {H.~A.}\ \bibnamefont {Thiessen}}, \bibinfo {author}
  {\bibfnamefont {S.~J.}\ \bibnamefont {Greene}}, \bibinfo {author}
  {\bibfnamefont {W.~J.}\ \bibnamefont {Braithwaite}}, \bibinfo {author}
  {\bibfnamefont {W.~B.}\ \bibnamefont {Cottingame}}, \bibinfo {author}
  {\bibfnamefont {D.~B.}\ \bibnamefont {Holtkamp}}, \bibinfo {author}
  {\bibfnamefont {I.~B.}\ \bibnamefont {Moore}}, \ and\ \bibinfo {author}
  {\bibfnamefont {C.~F.}\ \bibnamefont {Moore}},\ }\href {\doibase
  10.1103/PhysRevC.22.1180} {\bibfield  {journal} {\bibinfo  {journal} {Phys.
  Rev. C}\ }\textbf {\bibinfo {volume} {22}},\ \bibinfo {pages} {1180}
  (\bibinfo {year} {1980})}\BibitemShut {NoStop}%
\bibitem [{\citenamefont {F\"ohl}\ \emph {et~al.}(1997)\citenamefont {F\"ohl},
  \citenamefont {Bilger}, \citenamefont {Clement}, \citenamefont {Gr\"ater},
  \citenamefont {Meier}, \citenamefont {P\"atzold}, \citenamefont {Schapler},
  \citenamefont {Wagner}, \citenamefont {Wilhelm}, \citenamefont {Kluge},
  \citenamefont {Wieser}, \citenamefont {Schepkin}, \citenamefont {Abela},
  \citenamefont {Foroughi},\ and\ \citenamefont {Renker}}]{Fohl:1997}%
  \BibitemOpen
  \bibfield  {author} {\bibinfo {author} {\bibfnamefont {K.}~\bibnamefont
  {F\"ohl}}, \bibinfo {author} {\bibfnamefont {R.}~\bibnamefont {Bilger}},
  \bibinfo {author} {\bibfnamefont {H.}~\bibnamefont {Clement}}, \bibinfo
  {author} {\bibfnamefont {J.}~\bibnamefont {Gr\"ater}}, \bibinfo {author}
  {\bibfnamefont {R.}~\bibnamefont {Meier}}, \bibinfo {author} {\bibfnamefont
  {J.}~\bibnamefont {P\"atzold}}, \bibinfo {author} {\bibfnamefont
  {D.}~\bibnamefont {Schapler}}, \bibinfo {author} {\bibfnamefont {G.~J.}\
  \bibnamefont {Wagner}}, \bibinfo {author} {\bibfnamefont {O.}~\bibnamefont
  {Wilhelm}}, \bibinfo {author} {\bibfnamefont {W.}~\bibnamefont {Kluge}},
  \bibinfo {author} {\bibfnamefont {R.}~\bibnamefont {Wieser}}, \bibinfo
  {author} {\bibfnamefont {M.}~\bibnamefont {Schepkin}}, \bibinfo {author}
  {\bibfnamefont {R.}~\bibnamefont {Abela}}, \bibinfo {author} {\bibfnamefont
  {F.}~\bibnamefont {Foroughi}}, \ and\ \bibinfo {author} {\bibfnamefont
  {D.}~\bibnamefont {Renker}},\ }\href {\doibase 10.1103/PhysRevLett.79.3849}
  {\bibfield  {journal} {\bibinfo  {journal} {Phys. Rev. Lett.}\ }\textbf
  {\bibinfo {volume} {79}},\ \bibinfo {pages} {3849} (\bibinfo {year}
  {1997})}\BibitemShut {NoStop}%
\bibitem [{\citenamefont {Mukha}\ \emph {et~al.}(2010)\citenamefont {Mukha},
  \citenamefont {S\"ummerer}, \citenamefont {Acosta}, \citenamefont {Alvarez},
  \citenamefont {Casarejos}, \citenamefont {Chatillon}, \citenamefont
  {Cortina-Gil}, \citenamefont {Egorova}, \citenamefont {Espino}, \citenamefont
  {Fomichev}, \citenamefont {Garc\'ia-Ramos}, \citenamefont {Geissel},
  \citenamefont {G\'omez-Camacho}, \citenamefont {Grigorenko}, \citenamefont
  {Hofmann}, \citenamefont {Kiselev}, \citenamefont {Korsheninnikov},
  \citenamefont {Kurz}, \citenamefont {Litvinov}, \citenamefont {Litvinova},
  \citenamefont {Martel}, \citenamefont {Nociforo}, \citenamefont {Ott},
  \citenamefont {Pf\"utzner}, \citenamefont {Rodr\'iguez-Tajes}, \citenamefont
  {Roeckl}, \citenamefont {Stanoiu}, \citenamefont {Timofeyuk}, \citenamefont
  {Weick},\ and\ \citenamefont {Woods}}]{Mukha:2010}%
  \BibitemOpen
  \bibfield  {author} {\bibinfo {author} {\bibfnamefont {I.}~\bibnamefont
  {Mukha}}, \bibinfo {author} {\bibfnamefont {K.}~\bibnamefont {S\"ummerer}},
  \bibinfo {author} {\bibfnamefont {L.}~\bibnamefont {Acosta}}, \bibinfo
  {author} {\bibfnamefont {M.~A.~G.}\ \bibnamefont {Alvarez}}, \bibinfo
  {author} {\bibfnamefont {E.}~\bibnamefont {Casarejos}}, \bibinfo {author}
  {\bibfnamefont {A.}~\bibnamefont {Chatillon}}, \bibinfo {author}
  {\bibfnamefont {D.}~\bibnamefont {Cortina-Gil}}, \bibinfo {author}
  {\bibfnamefont {I.~A.}\ \bibnamefont {Egorova}}, \bibinfo {author}
  {\bibfnamefont {J.~M.}\ \bibnamefont {Espino}}, \bibinfo {author}
  {\bibfnamefont {A.}~\bibnamefont {Fomichev}}, \bibinfo {author}
  {\bibfnamefont {J.~E.}\ \bibnamefont {Garc\'ia-Ramos}}, \bibinfo {author}
  {\bibfnamefont {H.}~\bibnamefont {Geissel}}, \bibinfo {author} {\bibfnamefont
  {J.}~\bibnamefont {G\'omez-Camacho}}, \bibinfo {author} {\bibfnamefont
  {L.}~\bibnamefont {Grigorenko}}, \bibinfo {author} {\bibfnamefont
  {J.}~\bibnamefont {Hofmann}}, \bibinfo {author} {\bibfnamefont
  {O.}~\bibnamefont {Kiselev}}, \bibinfo {author} {\bibfnamefont
  {A.}~\bibnamefont {Korsheninnikov}}, \bibinfo {author} {\bibfnamefont
  {N.}~\bibnamefont {Kurz}}, \bibinfo {author} {\bibfnamefont {Y.~A.}\
  \bibnamefont {Litvinov}}, \bibinfo {author} {\bibfnamefont {E.}~\bibnamefont
  {Litvinova}}, \bibinfo {author} {\bibfnamefont {I.}~\bibnamefont {Martel}},
  \bibinfo {author} {\bibfnamefont {C.}~\bibnamefont {Nociforo}}, \bibinfo
  {author} {\bibfnamefont {W.}~\bibnamefont {Ott}}, \bibinfo {author}
  {\bibfnamefont {M.}~\bibnamefont {Pf\"utzner}}, \bibinfo {author}
  {\bibfnamefont {C.}~\bibnamefont {Rodr\'iguez-Tajes}}, \bibinfo {author}
  {\bibfnamefont {E.}~\bibnamefont {Roeckl}}, \bibinfo {author} {\bibfnamefont
  {M.}~\bibnamefont {Stanoiu}}, \bibinfo {author} {\bibfnamefont {N.~K.}\
  \bibnamefont {Timofeyuk}}, \bibinfo {author} {\bibfnamefont {H.}~\bibnamefont
  {Weick}}, \ and\ \bibinfo {author} {\bibfnamefont {P.~J.}\ \bibnamefont
  {Woods}},\ }\href {\doibase 10.1103/PhysRevC.82.054315} {\bibfield  {journal}
  {\bibinfo  {journal} {Phys. Rev. C}\ }\textbf {\bibinfo {volume} {82}},\
  \bibinfo {pages} {054315} (\bibinfo {year} {2010})}\BibitemShut {NoStop}%
\bibitem [{\citenamefont {Wamers}\ \emph {et~al.}(2014)\citenamefont {Wamers},
  \citenamefont {Marganiec}, \citenamefont {Aksouh}, \citenamefont {Aksyutina},
  \citenamefont {\'Alvarez-Pol}, \citenamefont {Aumann}, \citenamefont
  {Beceiro-Novo}, \citenamefont {Boretzky}, \citenamefont {Borge},
  \citenamefont {Chartier}, \citenamefont {Chatillon}, \citenamefont {Chulkov},
  \citenamefont {Cortina-Gil}, \citenamefont {Emling}, \citenamefont {Ershova},
  \citenamefont {Fraile}, \citenamefont {Fynbo}, \citenamefont {Galaviz},
  \citenamefont {Geissel}, \citenamefont {Heil}, \citenamefont {Hoffmann},
  \citenamefont {Johansson}, \citenamefont {Jonson}, \citenamefont
  {Karagiannis}, \citenamefont {Kiselev}, \citenamefont {Kratz}, \citenamefont
  {Kulessa}, \citenamefont {Kurz}, \citenamefont {Langer}, \citenamefont
  {Lantz}, \citenamefont {Le~Bleis}, \citenamefont {Lemmon}, \citenamefont
  {Litvinov}, \citenamefont {Mahata}, \citenamefont {M\"untz}, \citenamefont
  {Nilsson}, \citenamefont {Nociforo}, \citenamefont {Nyman}, \citenamefont
  {Ott}, \citenamefont {Panin}, \citenamefont {Paschalis}, \citenamefont
  {Perea}, \citenamefont {Plag}, \citenamefont {Reifarth}, \citenamefont
  {Richter}, \citenamefont {Rodriguez-Tajes}, \citenamefont {Rossi},
  \citenamefont {Riisager}, \citenamefont {Savran}, \citenamefont {Schrieder},
  \citenamefont {Simon}, \citenamefont {Stroth}, \citenamefont {S\"ummerer},
  \citenamefont {Tengblad}, \citenamefont {Weick}, \citenamefont {Wimmer},\
  and\ \citenamefont {Zhukov}}]{Wamers:2014}%
  \BibitemOpen
  \bibfield  {author} {\bibinfo {author} {\bibfnamefont {F.}~\bibnamefont
  {Wamers}}, \bibinfo {author} {\bibfnamefont {J.}~\bibnamefont {Marganiec}},
  \bibinfo {author} {\bibfnamefont {F.}~\bibnamefont {Aksouh}}, \bibinfo
  {author} {\bibfnamefont {Y.}~\bibnamefont {Aksyutina}}, \bibinfo {author}
  {\bibfnamefont {H.}~\bibnamefont {\'Alvarez-Pol}}, \bibinfo {author}
  {\bibfnamefont {T.}~\bibnamefont {Aumann}}, \bibinfo {author} {\bibfnamefont
  {S.}~\bibnamefont {Beceiro-Novo}}, \bibinfo {author} {\bibfnamefont
  {K.}~\bibnamefont {Boretzky}}, \bibinfo {author} {\bibfnamefont {M.~J.~G.}\
  \bibnamefont {Borge}}, \bibinfo {author} {\bibfnamefont {M.}~\bibnamefont
  {Chartier}}, \bibinfo {author} {\bibfnamefont {A.}~\bibnamefont {Chatillon}},
  \bibinfo {author} {\bibfnamefont {L.~V.}\ \bibnamefont {Chulkov}}, \bibinfo
  {author} {\bibfnamefont {D.}~\bibnamefont {Cortina-Gil}}, \bibinfo {author}
  {\bibfnamefont {H.}~\bibnamefont {Emling}}, \bibinfo {author} {\bibfnamefont
  {O.}~\bibnamefont {Ershova}}, \bibinfo {author} {\bibfnamefont {L.~M.}\
  \bibnamefont {Fraile}}, \bibinfo {author} {\bibfnamefont {H.~O.~U.}\
  \bibnamefont {Fynbo}}, \bibinfo {author} {\bibfnamefont {D.}~\bibnamefont
  {Galaviz}}, \bibinfo {author} {\bibfnamefont {H.}~\bibnamefont {Geissel}},
  \bibinfo {author} {\bibfnamefont {M.}~\bibnamefont {Heil}}, \bibinfo {author}
  {\bibfnamefont {D.~H.~H.}\ \bibnamefont {Hoffmann}}, \bibinfo {author}
  {\bibfnamefont {H.~T.}\ \bibnamefont {Johansson}}, \bibinfo {author}
  {\bibfnamefont {B.}~\bibnamefont {Jonson}}, \bibinfo {author} {\bibfnamefont
  {C.}~\bibnamefont {Karagiannis}}, \bibinfo {author} {\bibfnamefont {O.~A.}\
  \bibnamefont {Kiselev}}, \bibinfo {author} {\bibfnamefont {J.~V.}\
  \bibnamefont {Kratz}}, \bibinfo {author} {\bibfnamefont {R.}~\bibnamefont
  {Kulessa}}, \bibinfo {author} {\bibfnamefont {N.}~\bibnamefont {Kurz}},
  \bibinfo {author} {\bibfnamefont {C.}~\bibnamefont {Langer}}, \bibinfo
  {author} {\bibfnamefont {M.}~\bibnamefont {Lantz}}, \bibinfo {author}
  {\bibfnamefont {T.}~\bibnamefont {Le~Bleis}}, \bibinfo {author}
  {\bibfnamefont {R.}~\bibnamefont {Lemmon}}, \bibinfo {author} {\bibfnamefont
  {Y.~A.}\ \bibnamefont {Litvinov}}, \bibinfo {author} {\bibfnamefont
  {K.}~\bibnamefont {Mahata}}, \bibinfo {author} {\bibfnamefont
  {C.}~\bibnamefont {M\"untz}}, \bibinfo {author} {\bibfnamefont
  {T.}~\bibnamefont {Nilsson}}, \bibinfo {author} {\bibfnamefont
  {C.}~\bibnamefont {Nociforo}}, \bibinfo {author} {\bibfnamefont
  {G.}~\bibnamefont {Nyman}}, \bibinfo {author} {\bibfnamefont
  {W.}~\bibnamefont {Ott}}, \bibinfo {author} {\bibfnamefont {V.}~\bibnamefont
  {Panin}}, \bibinfo {author} {\bibfnamefont {S.}~\bibnamefont {Paschalis}},
  \bibinfo {author} {\bibfnamefont {A.}~\bibnamefont {Perea}}, \bibinfo
  {author} {\bibfnamefont {R.}~\bibnamefont {Plag}}, \bibinfo {author}
  {\bibfnamefont {R.}~\bibnamefont {Reifarth}}, \bibinfo {author}
  {\bibfnamefont {A.}~\bibnamefont {Richter}}, \bibinfo {author} {\bibfnamefont
  {C.}~\bibnamefont {Rodriguez-Tajes}}, \bibinfo {author} {\bibfnamefont
  {D.}~\bibnamefont {Rossi}}, \bibinfo {author} {\bibfnamefont
  {K.}~\bibnamefont {Riisager}}, \bibinfo {author} {\bibfnamefont
  {D.}~\bibnamefont {Savran}}, \bibinfo {author} {\bibfnamefont
  {G.}~\bibnamefont {Schrieder}}, \bibinfo {author} {\bibfnamefont
  {H.}~\bibnamefont {Simon}}, \bibinfo {author} {\bibfnamefont
  {J.}~\bibnamefont {Stroth}}, \bibinfo {author} {\bibfnamefont
  {K.}~\bibnamefont {S\"ummerer}}, \bibinfo {author} {\bibfnamefont
  {O.}~\bibnamefont {Tengblad}}, \bibinfo {author} {\bibfnamefont
  {H.}~\bibnamefont {Weick}}, \bibinfo {author} {\bibfnamefont
  {C.}~\bibnamefont {Wimmer}}, \ and\ \bibinfo {author} {\bibfnamefont {M.~V.}\
  \bibnamefont {Zhukov}},\ }\href {\doibase 10.1103/PhysRevLett.112.132502}
  {\bibfield  {journal} {\bibinfo  {journal} {Phys. Rev. Lett.}\ }\textbf
  {\bibinfo {volume} {112}},\ \bibinfo {pages} {132502} (\bibinfo {year}
  {2014})}\BibitemShut {NoStop}%
\bibitem [{\citenamefont {Kikuchi}\ \emph {et~al.}(2013)\citenamefont
  {Kikuchi}, \citenamefont {Matsumoto}, \citenamefont {Minomo},\ and\
  \citenamefont {Ogata}}]{Kikuchi:2013}%
  \BibitemOpen
  \bibfield  {author} {\bibinfo {author} {\bibfnamefont {Y.}~\bibnamefont
  {Kikuchi}}, \bibinfo {author} {\bibfnamefont {T.}~\bibnamefont {Matsumoto}},
  \bibinfo {author} {\bibfnamefont {K.}~\bibnamefont {Minomo}}, \ and\ \bibinfo
  {author} {\bibfnamefont {K.}~\bibnamefont {Ogata}},\ }\href {\doibase
  10.1103/PhysRevC.88.021602} {\bibfield  {journal} {\bibinfo  {journal} {Phys.
  Rev. C}\ }\textbf {\bibinfo {volume} {88}},\ \bibinfo {pages} {021602}
  (\bibinfo {year} {2013})}\BibitemShut {NoStop}%
\bibitem [{\citenamefont {Grigorenko}\ \emph {et~al.}(2013)\citenamefont
  {Grigorenko}, \citenamefont {Mukha},\ and\ \citenamefont
  {Zhukov}}]{Grigorenko:2013}%
  \BibitemOpen
  \bibfield  {author} {\bibinfo {author} {\bibfnamefont {L.~V.}\ \bibnamefont
  {Grigorenko}}, \bibinfo {author} {\bibfnamefont {I.~G.}\ \bibnamefont
  {Mukha}}, \ and\ \bibinfo {author} {\bibfnamefont {M.~V.}\ \bibnamefont
  {Zhukov}},\ }\href {\doibase 10.1103/PhysRevLett.111.042501} {\bibfield
  {journal} {\bibinfo  {journal} {Phys. Rev. Lett.}\ }\textbf {\bibinfo
  {volume} {111}},\ \bibinfo {pages} {042501} (\bibinfo {year}
  {2013})}\BibitemShut {NoStop}%
\bibitem [{\citenamefont {Hagino}\ and\ \citenamefont
  {Sagawa}(2014)}]{Hagino:2014}%
  \BibitemOpen
  \bibfield  {author} {\bibinfo {author} {\bibfnamefont {K.}~\bibnamefont
  {Hagino}}\ and\ \bibinfo {author} {\bibfnamefont {H.}~\bibnamefont
  {Sagawa}},\ }\href {\doibase 10.1103/PhysRevC.89.014331} {\bibfield
  {journal} {\bibinfo  {journal} {Phys. Rev. C}\ }\textbf {\bibinfo {volume}
  {89}},\ \bibinfo {pages} {014331} (\bibinfo {year} {2014})}\BibitemShut
  {NoStop}%
\bibitem [{\citenamefont {Wallace}\ \emph {et~al.}(2007)\citenamefont
  {Wallace}, \citenamefont {Famiano}, \citenamefont {van Goethem},
  \citenamefont {Rogers}, \citenamefont {Lynch}, \citenamefont {Clifford},
  \citenamefont {Delaunay}, \citenamefont {Lee}, \citenamefont {Labostov},
  \citenamefont {Mocko}, \citenamefont {Morris}, \citenamefont {Moroni},
  \citenamefont {Nett}, \citenamefont {Oostdyk}, \citenamefont {Krishnasamy},
  \citenamefont {Tsang}, \citenamefont {de~Souza}, \citenamefont {Hudan},
  \citenamefont {Sobotka}, \citenamefont {Charity}, \citenamefont {Elson},\
  and\ \citenamefont {Engel}}]{Wallace:2007}%
  \BibitemOpen
  \bibfield  {author} {\bibinfo {author} {\bibfnamefont {M.}~\bibnamefont
  {Wallace}}, \bibinfo {author} {\bibfnamefont {M.}~\bibnamefont {Famiano}},
  \bibinfo {author} {\bibfnamefont {M.-J.}\ \bibnamefont {van Goethem}},
  \bibinfo {author} {\bibfnamefont {A.}~\bibnamefont {Rogers}}, \bibinfo
  {author} {\bibfnamefont {W.}~\bibnamefont {Lynch}}, \bibinfo {author}
  {\bibfnamefont {J.}~\bibnamefont {Clifford}}, \bibinfo {author}
  {\bibfnamefont {F.}~\bibnamefont {Delaunay}}, \bibinfo {author}
  {\bibfnamefont {J.}~\bibnamefont {Lee}}, \bibinfo {author} {\bibfnamefont
  {S.}~\bibnamefont {Labostov}}, \bibinfo {author} {\bibfnamefont
  {M.}~\bibnamefont {Mocko}}, \bibinfo {author} {\bibfnamefont
  {L.}~\bibnamefont {Morris}}, \bibinfo {author} {\bibfnamefont
  {A.}~\bibnamefont {Moroni}}, \bibinfo {author} {\bibfnamefont
  {B.}~\bibnamefont {Nett}}, \bibinfo {author} {\bibfnamefont {D.}~\bibnamefont
  {Oostdyk}}, \bibinfo {author} {\bibfnamefont {R.}~\bibnamefont
  {Krishnasamy}}, \bibinfo {author} {\bibfnamefont {M.}~\bibnamefont {Tsang}},
  \bibinfo {author} {\bibfnamefont {R.}~\bibnamefont {de~Souza}}, \bibinfo
  {author} {\bibfnamefont {S.}~\bibnamefont {Hudan}}, \bibinfo {author}
  {\bibfnamefont {L.}~\bibnamefont {Sobotka}}, \bibinfo {author} {\bibfnamefont
  {R.}~\bibnamefont {Charity}}, \bibinfo {author} {\bibfnamefont
  {J.}~\bibnamefont {Elson}}, \ and\ \bibinfo {author} {\bibfnamefont
  {G.}~\bibnamefont {Engel}},\ }\href {\doibase
  http://dx.doi.org/10.1016/j.nima.2007.08.248} {\bibfield  {journal} {\bibinfo
   {journal} {Nuclear Instruments and Methods in Physics Research Section A:
  Accelerators, Spectrometers, Detectors and Associated Equipment}\ }\textbf
  {\bibinfo {volume} {583}},\ \bibinfo {pages} {302 } (\bibinfo {year}
  {2007})}\BibitemShut {NoStop}%
\bibitem [{\citenamefont {Charity}\ \emph {et~al.}(2011)\citenamefont
  {Charity}, \citenamefont {Elson}, \citenamefont {Manfredi}, \citenamefont
  {Shane}, \citenamefont {Sobotka}, \citenamefont {Brown}, \citenamefont
  {Chajecki}, \citenamefont {Coupland}, \citenamefont {Iwasaki}, \citenamefont
  {Kilburn}, \citenamefont {Lee}, \citenamefont {Lynch}, \citenamefont
  {Sanetullaev}, \citenamefont {Tsang}, \citenamefont {Winkelbauer},
  \citenamefont {Youngs}, \citenamefont {Marley}, \citenamefont {Shetty},
  \citenamefont {Wuosmaa}, \citenamefont {Ghosh},\ and\ \citenamefont
  {Howard}}]{Charity:2011}%
  \BibitemOpen
  \bibfield  {author} {\bibinfo {author} {\bibfnamefont {R.~J.}\ \bibnamefont
  {Charity}}, \bibinfo {author} {\bibfnamefont {J.~M.}\ \bibnamefont {Elson}},
  \bibinfo {author} {\bibfnamefont {J.}~\bibnamefont {Manfredi}}, \bibinfo
  {author} {\bibfnamefont {R.}~\bibnamefont {Shane}}, \bibinfo {author}
  {\bibfnamefont {L.~G.}\ \bibnamefont {Sobotka}}, \bibinfo {author}
  {\bibfnamefont {B.~A.}\ \bibnamefont {Brown}}, \bibinfo {author}
  {\bibfnamefont {Z.}~\bibnamefont {Chajecki}}, \bibinfo {author}
  {\bibfnamefont {D.}~\bibnamefont {Coupland}}, \bibinfo {author}
  {\bibfnamefont {H.}~\bibnamefont {Iwasaki}}, \bibinfo {author} {\bibfnamefont
  {M.}~\bibnamefont {Kilburn}}, \bibinfo {author} {\bibfnamefont
  {J.}~\bibnamefont {Lee}}, \bibinfo {author} {\bibfnamefont {W.~G.}\
  \bibnamefont {Lynch}}, \bibinfo {author} {\bibfnamefont {A.}~\bibnamefont
  {Sanetullaev}}, \bibinfo {author} {\bibfnamefont {M.~B.}\ \bibnamefont
  {Tsang}}, \bibinfo {author} {\bibfnamefont {J.}~\bibnamefont {Winkelbauer}},
  \bibinfo {author} {\bibfnamefont {M.}~\bibnamefont {Youngs}}, \bibinfo
  {author} {\bibfnamefont {S.~T.}\ \bibnamefont {Marley}}, \bibinfo {author}
  {\bibfnamefont {D.~V.}\ \bibnamefont {Shetty}}, \bibinfo {author}
  {\bibfnamefont {A.~H.}\ \bibnamefont {Wuosmaa}}, \bibinfo {author}
  {\bibfnamefont {T.~K.}\ \bibnamefont {Ghosh}}, \ and\ \bibinfo {author}
  {\bibfnamefont {M.~E.}\ \bibnamefont {Howard}},\ }\href {\doibase
  10.1103/PhysRevC.84.014320} {\bibfield  {journal} {\bibinfo  {journal} {Phys.
  Rev. C}\ }\textbf {\bibinfo {volume} {84}},\ \bibinfo {pages} {014320}
  (\bibinfo {year} {2011})}\BibitemShut {NoStop}%
\bibitem [{\citenamefont {Grigorenko}\ \emph {et~al.}(2002)\citenamefont
  {Grigorenko}, \citenamefont {Mukha}, \citenamefont {Thompson},\ and\
  \citenamefont {Zhukov}}]{Grigorenko:2002}%
  \BibitemOpen
  \bibfield  {author} {\bibinfo {author} {\bibfnamefont {L.~V.}\ \bibnamefont
  {Grigorenko}}, \bibinfo {author} {\bibfnamefont {I.~G.}\ \bibnamefont
  {Mukha}}, \bibinfo {author} {\bibfnamefont {I.~J.}\ \bibnamefont {Thompson}},
  \ and\ \bibinfo {author} {\bibfnamefont {M.~V.}\ \bibnamefont {Zhukov}},\
  }\href {\doibase 10.1103/PhysRevLett.88.042502} {\bibfield  {journal}
  {\bibinfo  {journal} {Phys. Rev. Lett.}\ }\textbf {\bibinfo {volume} {88}},\
  \bibinfo {pages} {042502} (\bibinfo {year} {2002})}\BibitemShut {NoStop}%
\bibitem [{\citenamefont {Grigorenko}\ \emph {et~al.}(2009)\citenamefont
  {Grigorenko}, \citenamefont {Wiser}, \citenamefont {Mercurio}, \citenamefont
  {Charity}, \citenamefont {Shane}, \citenamefont {Sobotka}, \citenamefont
  {Elson}, \citenamefont {Wuosmaa}, \citenamefont {Banu}, \citenamefont
  {McCleskey}, \citenamefont {Trache}, \citenamefont {Tribble},\ and\
  \citenamefont {Zhukov}}]{Grigorenko:2009c}%
  \BibitemOpen
  \bibfield  {author} {\bibinfo {author} {\bibfnamefont {L.~V.}\ \bibnamefont
  {Grigorenko}}, \bibinfo {author} {\bibfnamefont {T.~D.}\ \bibnamefont
  {Wiser}}, \bibinfo {author} {\bibfnamefont {K.}~\bibnamefont {Mercurio}},
  \bibinfo {author} {\bibfnamefont {R.~J.}\ \bibnamefont {Charity}}, \bibinfo
  {author} {\bibfnamefont {R.}~\bibnamefont {Shane}}, \bibinfo {author}
  {\bibfnamefont {L.~G.}\ \bibnamefont {Sobotka}}, \bibinfo {author}
  {\bibfnamefont {J.~M.}\ \bibnamefont {Elson}}, \bibinfo {author}
  {\bibfnamefont {A.~H.}\ \bibnamefont {Wuosmaa}}, \bibinfo {author}
  {\bibfnamefont {A.}~\bibnamefont {Banu}}, \bibinfo {author} {\bibfnamefont
  {M.}~\bibnamefont {McCleskey}}, \bibinfo {author} {\bibfnamefont
  {L.}~\bibnamefont {Trache}}, \bibinfo {author} {\bibfnamefont {R.~E.}\
  \bibnamefont {Tribble}}, \ and\ \bibinfo {author} {\bibfnamefont {M.~V.}\
  \bibnamefont {Zhukov}},\ }\href {\doibase 10.1103/PhysRevC.80.034602}
  {\bibfield  {journal} {\bibinfo  {journal} {Phys. Rev. C}\ }\textbf {\bibinfo
  {volume} {80}},\ \bibinfo {pages} {034602} (\bibinfo {year}
  {2009})}\BibitemShut {NoStop}%
\bibitem [{\citenamefont {Grigorenko}\ \emph {et~al.}(2005)\citenamefont
  {Grigorenko}, \citenamefont {Parfenova},\ and\ \citenamefont
  {Zhukov}}]{Grigorenko:2005}%
  \BibitemOpen
  \bibfield  {author} {\bibinfo {author} {\bibfnamefont {L.~V.}\ \bibnamefont
  {Grigorenko}}, \bibinfo {author} {\bibfnamefont {Y.~L.}\ \bibnamefont
  {Parfenova}}, \ and\ \bibinfo {author} {\bibfnamefont {M.~V.}\ \bibnamefont
  {Zhukov}},\ }\href {\doibase 10.1103/PhysRevC.71.051604} {\bibfield
  {journal} {\bibinfo  {journal} {Phys. Rev. C}\ }\textbf {\bibinfo {volume}
  {71}},\ \bibinfo {pages} {051604} (\bibinfo {year} {2005})}\BibitemShut
  {NoStop}%
\bibitem [{\citenamefont {Goldberg}\ \emph {et~al.}(2004)\citenamefont
  {Goldberg}, \citenamefont {Chubarian}, \citenamefont {Tabacaru},
  \citenamefont {Trache}, \citenamefont {Tribble}, \citenamefont {Aprahamian},
  \citenamefont {Rogachev}, \citenamefont {Skorodumov},\ and\ \citenamefont
  {Tang}}]{Goldberg:2004}%
  \BibitemOpen
  \bibfield  {author} {\bibinfo {author} {\bibfnamefont {V.~Z.}\ \bibnamefont
  {Goldberg}}, \bibinfo {author} {\bibfnamefont {G.~G.}\ \bibnamefont
  {Chubarian}}, \bibinfo {author} {\bibfnamefont {G.}~\bibnamefont {Tabacaru}},
  \bibinfo {author} {\bibfnamefont {L.}~\bibnamefont {Trache}}, \bibinfo
  {author} {\bibfnamefont {R.~E.}\ \bibnamefont {Tribble}}, \bibinfo {author}
  {\bibfnamefont {A.}~\bibnamefont {Aprahamian}}, \bibinfo {author}
  {\bibfnamefont {G.~V.}\ \bibnamefont {Rogachev}}, \bibinfo {author}
  {\bibfnamefont {B.~B.}\ \bibnamefont {Skorodumov}}, \ and\ \bibinfo {author}
  {\bibfnamefont {X.~D.}\ \bibnamefont {Tang}},\ }\href {\doibase
  10.1103/PhysRevC.69.031302} {\bibfield  {journal} {\bibinfo  {journal} {Phys.
  Rev. C}\ }\textbf {\bibinfo {volume} {69}},\ \bibinfo {pages} {031302}
  (\bibinfo {year} {2004})}\BibitemShut {NoStop}%
\bibitem [{\citenamefont {Gogny}\ \emph {et~al.}(1970)\citenamefont {Gogny},
  \citenamefont {Pires},\ and\ \citenamefont {Tourreil}}]{Gogny:1970}%
  \BibitemOpen
  \bibfield  {author} {\bibinfo {author} {\bibfnamefont {D.}~\bibnamefont
  {Gogny}}, \bibinfo {author} {\bibfnamefont {P.}~\bibnamefont {Pires}}, \ and\
  \bibinfo {author} {\bibfnamefont {R.~D.}\ \bibnamefont {Tourreil}},\ }\href
  {\doibase http://dx.doi.org/10.1016/0370-2693(70)90552-6} {\bibfield
  {journal} {\bibinfo  {journal} {Physics Letters B}\ }\textbf {\bibinfo
  {volume} {32}},\ \bibinfo {pages} {591 } (\bibinfo {year}
  {1970})}\BibitemShut {NoStop}%
\bibitem [{\citenamefont {Grigorenko}\ \emph {et~al.}(2001)\citenamefont
  {Grigorenko}, \citenamefont {Johnson}, \citenamefont {Mukha}, \citenamefont
  {Thompson},\ and\ \citenamefont {Zhukov}}]{Grigorenko:2001}%
  \BibitemOpen
  \bibfield  {author} {\bibinfo {author} {\bibfnamefont {L.~V.}\ \bibnamefont
  {Grigorenko}}, \bibinfo {author} {\bibfnamefont {R.~C.}\ \bibnamefont
  {Johnson}}, \bibinfo {author} {\bibfnamefont {I.~G.}\ \bibnamefont {Mukha}},
  \bibinfo {author} {\bibfnamefont {I.~J.}\ \bibnamefont {Thompson}}, \ and\
  \bibinfo {author} {\bibfnamefont {M.~V.}\ \bibnamefont {Zhukov}},\ }\href
  {\doibase 10.1103/PhysRevC.64.054002} {\bibfield  {journal} {\bibinfo
  {journal} {Phys. Rev. C}\ }\textbf {\bibinfo {volume} {64}},\ \bibinfo
  {pages} {054002} (\bibinfo {year} {2001})}\BibitemShut {NoStop}%
\bibitem [{\citenamefont {Grigorenko}\ and\ \citenamefont
  {Zhukov}(2007)}]{Grigorenko:2007}%
  \BibitemOpen
  \bibfield  {author} {\bibinfo {author} {\bibfnamefont {L.~V.}\ \bibnamefont
  {Grigorenko}}\ and\ \bibinfo {author} {\bibfnamefont {M.~V.}\ \bibnamefont
  {Zhukov}},\ }\href {\doibase 10.1103/PhysRevC.76.014008} {\bibfield
  {journal} {\bibinfo  {journal} {Phys. Rev. C}\ }\textbf {\bibinfo {volume}
  {76}},\ \bibinfo {pages} {014008} (\bibinfo {year} {2007})}\BibitemShut
  {NoStop}%
\bibitem [{\citenamefont {Korsheninnikov}(1990)}]{Korsheninnikov:1990}%
  \BibitemOpen
  \bibfield  {author} {\bibinfo {author} {\bibfnamefont {A.~A.}\ \bibnamefont
  {Korsheninnikov}},\ }\href@noop {} {\bibfield  {journal} {\bibinfo  {journal}
  {Sov. J. Nucl. Phys. (Yad. Fiz. )}\ }\textbf {\bibinfo {volume} {52}},\
  \bibinfo {pages} {827} (\bibinfo {year} {1990})}\BibitemShut {NoStop}%
\bibitem [{\citenamefont {Azhari}\ \emph {et~al.}(1998)\citenamefont {Azhari},
  \citenamefont {Kryger},\ and\ \citenamefont {Thoennessen}}]{Azhari:1998}%
  \BibitemOpen
  \bibfield  {author} {\bibinfo {author} {\bibfnamefont {A.}~\bibnamefont
  {Azhari}}, \bibinfo {author} {\bibfnamefont {R.~A.}\ \bibnamefont {Kryger}},
  \ and\ \bibinfo {author} {\bibfnamefont {M.}~\bibnamefont {Thoennessen}},\
  }\href {\doibase 10.1103/PhysRevC.58.2568} {\bibfield  {journal} {\bibinfo
  {journal} {Phys. Rev. C}\ }\textbf {\bibinfo {volume} {58}},\ \bibinfo
  {pages} {2568} (\bibinfo {year} {1998})}\BibitemShut {NoStop}%
\bibitem [{\citenamefont {Barker}(2003)}]{Barker:2003}%
  \BibitemOpen
  \bibfield  {author} {\bibinfo {author} {\bibfnamefont {F.~C.}\ \bibnamefont
  {Barker}},\ }\href {\doibase 10.1103/PhysRevC.68.054602} {\bibfield
  {journal} {\bibinfo  {journal} {Phys. Rev. C}\ }\textbf {\bibinfo {volume}
  {68}},\ \bibinfo {pages} {054602} (\bibinfo {year} {2003})}\BibitemShut
  {NoStop}%
\bibitem [{\citenamefont {Fortune}\ and\ \citenamefont
  {Sherr}(2003)}]{Fortune:2003}%
  \BibitemOpen
  \bibfield  {author} {\bibinfo {author} {\bibfnamefont {H.~T.}\ \bibnamefont
  {Fortune}}\ and\ \bibinfo {author} {\bibfnamefont {R.}~\bibnamefont
  {Sherr}},\ }\href {\doibase 10.1103/PhysRevC.68.034309} {\bibfield  {journal}
  {\bibinfo  {journal} {Phys. Rev. C}\ }\textbf {\bibinfo {volume} {68}},\
  \bibinfo {pages} {034309} (\bibinfo {year} {2003})}\BibitemShut {NoStop}%
\bibitem [{\citenamefont {Jager}\ \emph {et~al.}(2012)\citenamefont {Jager},
  \citenamefont {Charity}, \citenamefont {Elson}, \citenamefont {Manfredi},
  \citenamefont {Mahzoon}, \citenamefont {Sobotka}, \citenamefont {McCleskey},
  \citenamefont {Pizzone}, \citenamefont {Roeder}, \citenamefont {Spiridon},
  \citenamefont {Simmons}, \citenamefont {Trache},\ and\ \citenamefont
  {Kurokawa}}]{Jager:2012}%
  \BibitemOpen
  \bibfield  {author} {\bibinfo {author} {\bibfnamefont {M.~F.}\ \bibnamefont
  {Jager}}, \bibinfo {author} {\bibfnamefont {R.~J.}\ \bibnamefont {Charity}},
  \bibinfo {author} {\bibfnamefont {J.~M.}\ \bibnamefont {Elson}}, \bibinfo
  {author} {\bibfnamefont {J.}~\bibnamefont {Manfredi}}, \bibinfo {author}
  {\bibfnamefont {M.~H.}\ \bibnamefont {Mahzoon}}, \bibinfo {author}
  {\bibfnamefont {L.~G.}\ \bibnamefont {Sobotka}}, \bibinfo {author}
  {\bibfnamefont {M.}~\bibnamefont {McCleskey}}, \bibinfo {author}
  {\bibfnamefont {R.~G.}\ \bibnamefont {Pizzone}}, \bibinfo {author}
  {\bibfnamefont {B.~T.}\ \bibnamefont {Roeder}}, \bibinfo {author}
  {\bibfnamefont {A.}~\bibnamefont {Spiridon}}, \bibinfo {author}
  {\bibfnamefont {E.}~\bibnamefont {Simmons}}, \bibinfo {author} {\bibfnamefont
  {L.}~\bibnamefont {Trache}}, \ and\ \bibinfo {author} {\bibfnamefont
  {M.}~\bibnamefont {Kurokawa}},\ }\href {\doibase 10.1103/PhysRevC.86.011304}
  {\bibfield  {journal} {\bibinfo  {journal} {Phys. Rev. C}\ }\textbf {\bibinfo
  {volume} {86}},\ \bibinfo {pages} {011304} (\bibinfo {year}
  {2012})}\BibitemShut {NoStop}%
\bibitem [{\citenamefont {Chromik}\ \emph {et~al.}(2002)\citenamefont
  {Chromik}, \citenamefont {Thirolf}, \citenamefont {Thoennessen},
  \citenamefont {Brown}, \citenamefont {Davinson}, \citenamefont {Gassmann},
  \citenamefont {Heckman}, \citenamefont {Prisciandaro}, \citenamefont
  {Reiter}, \citenamefont {Tryggestad},\ and\ \citenamefont
  {Woods}}]{Chromik:2002}%
  \BibitemOpen
  \bibfield  {author} {\bibinfo {author} {\bibfnamefont {M.~J.}\ \bibnamefont
  {Chromik}}, \bibinfo {author} {\bibfnamefont {P.~G.}\ \bibnamefont
  {Thirolf}}, \bibinfo {author} {\bibfnamefont {M.}~\bibnamefont
  {Thoennessen}}, \bibinfo {author} {\bibfnamefont {B.~A.}\ \bibnamefont
  {Brown}}, \bibinfo {author} {\bibfnamefont {T.}~\bibnamefont {Davinson}},
  \bibinfo {author} {\bibfnamefont {D.}~\bibnamefont {Gassmann}}, \bibinfo
  {author} {\bibfnamefont {P.}~\bibnamefont {Heckman}}, \bibinfo {author}
  {\bibfnamefont {J.}~\bibnamefont {Prisciandaro}}, \bibinfo {author}
  {\bibfnamefont {P.}~\bibnamefont {Reiter}}, \bibinfo {author} {\bibfnamefont
  {E.}~\bibnamefont {Tryggestad}}, \ and\ \bibinfo {author} {\bibfnamefont
  {P.~J.}\ \bibnamefont {Woods}},\ }\href {\doibase 10.1103/PhysRevC.66.024313}
  {\bibfield  {journal} {\bibinfo  {journal} {Phys. Rev. C}\ }\textbf {\bibinfo
  {volume} {66}},\ \bibinfo {pages} {024313} (\bibinfo {year}
  {2002})}\BibitemShut {NoStop}%
\bibitem [{\citenamefont {Fortune}(2006)}]{Fortune:2006}%
  \BibitemOpen
  \bibfield  {author} {\bibinfo {author} {\bibfnamefont {H.~T.}\ \bibnamefont
  {Fortune}},\ }\href {\doibase 10.1103/PhysRevC.74.054310} {\bibfield
  {journal} {\bibinfo  {journal} {Phys. Rev. C}\ }\textbf {\bibinfo {volume}
  {74}},\ \bibinfo {pages} {054310} (\bibinfo {year} {2006})}\BibitemShut
  {NoStop}%
\bibitem [{\citenamefont {Guo}\ \emph {et~al.}(2005)\citenamefont {Guo},
  \citenamefont {Powell}, \citenamefont {Lee}, \citenamefont {Leitner},
  \citenamefont {McMahan}, \citenamefont {Moltz}, \citenamefont {O'Neil},
  \citenamefont {Perajarvi}, \citenamefont {Phair}, \citenamefont {Ramsey},
  \citenamefont {Xu},\ and\ \citenamefont {Cerny}}]{Guo:2005}%
  \BibitemOpen
  \bibfield  {author} {\bibinfo {author} {\bibfnamefont {F.~Q.}\ \bibnamefont
  {Guo}}, \bibinfo {author} {\bibfnamefont {J.}~\bibnamefont {Powell}},
  \bibinfo {author} {\bibfnamefont {D.~W.}\ \bibnamefont {Lee}}, \bibinfo
  {author} {\bibfnamefont {D.}~\bibnamefont {Leitner}}, \bibinfo {author}
  {\bibfnamefont {M.~A.}\ \bibnamefont {McMahan}}, \bibinfo {author}
  {\bibfnamefont {D.~M.}\ \bibnamefont {Moltz}}, \bibinfo {author}
  {\bibfnamefont {J.~P.}\ \bibnamefont {O'Neil}}, \bibinfo {author}
  {\bibfnamefont {K.}~\bibnamefont {Perajarvi}}, \bibinfo {author}
  {\bibfnamefont {L.}~\bibnamefont {Phair}}, \bibinfo {author} {\bibfnamefont
  {C.~A.}\ \bibnamefont {Ramsey}}, \bibinfo {author} {\bibfnamefont {X.~J.}\
  \bibnamefont {Xu}}, \ and\ \bibinfo {author} {\bibfnamefont {J.}~\bibnamefont
  {Cerny}},\ }\href {\doibase 10.1103/PhysRevC.72.034312} {\bibfield  {journal}
  {\bibinfo  {journal} {Phys. Rev. C}\ }\textbf {\bibinfo {volume} {72}},\
  \bibinfo {pages} {034312} (\bibinfo {year} {2005})}\BibitemShut {NoStop}%
\end{thebibliography}
%


\end{document}